\documentclass[12pt]{article}
\usepackage{algorithm}
\usepackage{algpseudocode}
\usepackage{graphicx}
\usepackage{enumerate}
\usepackage{natbib}
\usepackage{url} 
\usepackage{amsmath}
\usepackage{mathrsfs}
\usepackage{mathtools}
\usepackage[shortlabels]{enumitem} 
\usepackage{float}
\usepackage{tikz}
\usepackage{placeins}
\usetikzlibrary{arrows,positioning,calc, external} 
\usepackage[graphicx]{realboxes}
\usepackage{amsthm}

\tikzset{
    >=stealth',
    punkt/.style={
           rectangle,
           rounded corners,
           draw=black, thick,
           text width=7em,
           minimum height=2em,
           text centered},
    punkts/.style={
           rectangle,
           rounded corners,
           draw=black, thick,
           text width=3em,
           minimum height=2em,
           text centered},       
    punktl/.style={
           re
           tangle,
           rounded corners,
           draw=black, thick,
           
           text width=7em,
           minimum height=2em,
           text centered},
    pil/.style={
           ->,
           shorten <=4pt,
           shorten >=4pt,},
    pildotted/.style={
           ->,
           shorten <=4pt,
           shorten >=4pt,
  dotted,
  },
  external/system call={pdflatex \tikzexternalcheckshellescape 
                                        -halt-on-error
                                        -interaction=batchmode 
                                        -jobname "\image" "\texsource"
                                        && pdftops -eps "\image.pdf"}
}

\usepackage{bm}

\usepackage{amsfonts}
\usepackage{graphicx}
\usepackage{xcolor}
\usepackage{amssymb}
\usepackage{prodint}
\usepackage{caption}
\usepackage[font=small]{caption}

\newcommand{\indep}{\perp \!\!\! \perp}
\newcommand*\diff{\mathop{}\!\mathrm{d}}

\newcommand{\Var}{\mathrm{Var}}

\newcommand\xqed[1]{%
  \leavevmode\unskip\penalty9999 \hbox{}\nobreak\hfill
  \quad\hbox{#1}}
\newcommand\demormk{\xqed{$\triangledown$}}

\newcommand\demoas{\xqed{$\diamond$}}

\makeatletter
\newcommand*{\bigcdot}{}
\DeclareRobustCommand*{\bigcdot}{%
  \mathbin{\mathpalette\bigcdot@{}}%
}
\newcommand*{\bigcdot@scalefactor}{.5}
\newcommand*{\bigcdot@widthfactor}{1.15}
\newcommand*{\bigcdot@}[2]{%
  \sbox0{$#1\vcenter{}$}
  \sbox2{$#1\cdot\m@th$}%
  \hbox to \bigcdot@widthfactor\wd2{%
    \hfil
    \raise\ht0\hbox{%
      \scalebox{\bigcdot@scalefactor}{%
        \lower\ht0\hbox{$#1\bullet\m@th$}%
      }%
    }%
    \hfil
  }%
}
\makeatother

\newtheorem{assumption}{Assumption}
\newtheorem{theorem}{Theorem}[section]
\newtheorem{corollary}{Corollary}[theorem]

\newtheorem{remark}{Remark}
\newtheorem{proposition}{Proposition}

\begin{document}

\def\spacingset#1{\renewcommand{\baselinestretch}%
{#1}\small\normalsize} \spacingset{1}


{
  \title{\bf A doubly robust learner for regression and inference with right-censored outcomes}
  \author{O.L.~Sandqvist$^{\textnormal{a},\textnormal{b}}$ \vspace{0.5cm}\\
    $^\textnormal{a}$PFA Pension, Sundkrogsgade 4, DK-2100 Copenhagen \O, Denmark \\
    $^\textnormal{b}$Department of Mathematical Sciences, University of Copenhagen, \\ Universitetsparken 5, DK-2100 Copenhagen \O, Denmark}
    
  \maketitle
}

\bigskip
\begin{abstract}
This paper introduces a general framework for doubly robust nonparametric regression with right-censored outcomes, adapting the strategy of the 'DR-learner' for heterogeneous treatment effects to the censored data setting. Our method generalizes censoring unbiased transformations to generate pseudo-outcomes, which serve as inputs for a second-stage nonparametric regression with sample-splitting or cross-fitting. We derive a novel representation of the conditional bias of these pseudo-outcomes; this representation is central to establishing the estimator’s asymptotic properties, including rate double robustness and oracle efficiency when the second-stage regression is a linear smoother. Simulation studies demonstrate the method’s favorable finite-sample performance, and we showcase its practical utility by applying it to estimate a causal effect within a regression discontinuity design with right-censored outcomes.
\end{abstract}

\noindent%
{\it Keywords:} Censoring unbiased transformation, machine learning, nonparametric regression, pseudo-outcomes, regression discontinuity design. 
\vfill

\newpage
\spacingset{1.2}
\section{Introduction} \label{sec:Introduction}

In many situations, one is interested in modeling the effect of covariates $W$ on an outcome $Y$. Powerful regression methods based on i.i.d. observations allow for flexible ways to estimate the conditional expectation function $\mathbb{E}[Y \mid W = w]$ without imposing strong parametric assumptions. However, when the outcome $Y$ is not fully observed due to censoring, standard regression methods applied to complete-case data can lead to substantial bias. This paper develops new methods for nonparametric regression in the presence of censoring, which is a type of missing data.

The problem of missing data in statistical inference has been extensively studied. One stream of literature, rooted in semiparametric efficiency theory, has developed powerful doubly robust and efficient-influence-function-based estimators. Seminal works, such as~\citet{Van:Robins:2003},~\citet{Chernozhukov.etal:2018}, as well as more recent cross-fitting-based methods such as~\citet{Rotnitzky.etal:2021},~\citet{Hou.etal:2023},~\citet{Wang.etal:2024}, and~\citet{Luo.etal:2025}, provide ways to obtain doubly robust estimates of low-dimensional scalar parameters, such as the average treatment effect or the parametric part of a Cox model. These methods are invaluable when the goal is to estimate a single summary measure of an effect. However, they do not directly apply to the estimation of infinite-dimensional objects like regression functions $\mathbb{E}[Y \mid W = w]$, which describe how the effect of covariates varies across the population. This is because when the estimand is not an expectation but rather a regression function, the efficient influence function of the estimand generally does not exist, and the classic theory on efficient estimation cannot be applied.

A stream of literature has emerged in causal inference to address the estimation of heterogeneous treatment effects. Methods like the DR-learner of~\citet{Kennedy:2023}, R-learner of~\citet{Nie:Wager:2021,Robinson:1988}, and X-learner of~\citet{Kuenzel:et:al:2019} have been proposed for estimation of the conditional average treatment effect (CATE). These approaches are powerful because they leverage machine learning to flexibly model the regression function while maintaining desirable statistical properties. However, these methods focus on missingness due to treatment assignment (i.e., not observing all counterfactual outcomes) and do not directly apply to the common problem of missingness due to censoring. The primary aim of this paper is to adapt the DR-learner methodology to the censored data setting. We refer to~\citet{Kennedy:2023} for a discussion of how the DR-learner compares to related literature such as~\citet{Ai:Chen:2003},~\citet{Rubin:Van:2005}, and~\citet{Foster:Syrgkanis:2023}.

The idea of the DR-learner is that semiparametrically efficient estimators estimate a marginal estimand $\mathbb{E}[Y]$ by averaging over the uncentered efficient influence function of $\mathbb{E}[Y]$, so a conditional estimand $\mathbb{E}[Y \mid W=w]$ may be estimated by regressing the uncentered efficient influence function of $\mathbb{E}[Y]$ on covariates. This approach fundamentally relies on constructing \textit{pseudo-outcomes}. Broadly speaking, pseudo-outcome methods are a statistical technique used to transform complex, partially unobserved data into fully observed proxy variables. These pseudo-outcomes are constructed so that a specific statistical property of the proxy -- most commonly its conditional expectation -- aligns with the estimand of interest from the unobserved full data. Because of this property, these proxies can be seamlessly plugged into standard, off-the-shelf regression models. One may recognize the doubly robust censoring unbiased transformation (DRCUT) pseudo-outcomes of~\citet{Rubin:Van:2007} and~\citet{Steingrimsson.etal:2016,Steingrimsson.etal:2019} as uncentered efficient influence functions, so they are in fact early proponents of an approach like the DR-learner and within the censored data domain no less. However, they only consider the survival setting, do not use cross-fitting, and do not provide any asymptotic theory. 

This paper extends the DRCUT and, in passing, also the regression discontinuity design (RDD) methodology to any right-censored data satisfying coarsening at random and positivity. A novel doubly robust representation of the conditional bias of the DRCUT is obtained. Using this representation, large sample properties of DRCUT-based estimators are established, including rate double robustness and oracle efficiency results when combined with cross-fitting and linear smoothers. While the overall proof strategy for establishing oracle efficiency is analogous, our analysis addresses the specific technical challenges that come with the censored data setting, namely the more complex conditional bias and nuisance functions. The latter consists of conditional hazards and Doob martingales, which may depend on the entire process history, rather than the propensity scores and conditional means typical of the CATE literature. Along the way, we extend the analysis of~\citet{Kennedy:2023} from sample-splitting to cross-fitting, and results on cross-fitting are extended to estimands that converge slower than $\sqrt{n}$. Simulation studies demonstrate favorable performance compared to existing methods. Finally, the approach is applied to the Longitudinal Study of Young People in England (LSYPE) in the form of an RDD to infer a CATE.

The paper is structured as follows. In Section~\ref{sec:DRTransform}, the DRCUT is proposed and a doubly robust representation for its conditional bias is presented. In Section~\ref{sec:Asymp}, the asymptotic properties of DRCUT-based estimators are studied, and it is shown that the approach achieves rate double robustness and model double robustness under suitable regularity assumptions. In Section~\ref{sec:Sim}, simulation experiments are conducted and Section~\ref{sec:Applic} contains the real data application. All implementations are available on GitHub at \url{https://github.com/oliversandqvist/Web-appendix-drcut}.

\section{Doubly robust censoring unbiased transformation} \label{sec:DRTransform}

 The full data (outcomes and covariates) is a càdlàg stochastic process $X=(X(t))_{t \geq 0}$ taking values in a metric space equipped with its Borel $\sigma$-algebra. Let $\mathcal{X}$ be the corresponding Skorokhod space. The process $X$ stopped at $u \geq 0$ is denoted $X^u = (X(u \wedge t))_{t \geq 0}$, where $\wedge$ returns the minimum of two values. The $(0,\infty)$-valued censoring variable is denoted $C$, so the observed data is $(C,X^C)$. It is assumed that $X(0)$ contains baseline covariates denoted by $W$. The outcome of interest is a real-valued function $Y(X)$, which is assumed to have finite second moment. Vector-valued outcomes can be accommodated by applying our results coordinate-wise, see Remark~\ref{rmk:VectorValued}. In practice, we can at most observe $X$ on a bounded interval $[0,\eta]$ for a fixed time horizon $\eta$, and we therefore assume $Y(X)=\tilde{Y}(X^\eta)$ for a suitable function $\Tilde{Y}$. If this does not hold, the approach in this paper fails since the DRCUT utilizes inverse probability of censoring weighting. In such cases, one might be able to use the proposed method for some outcome $\tilde{Y}(X^\eta)$ that approximates $Y(X)$, use other regression methods for the residual $Y(X)-\tilde{Y}(X^\eta)$, and then sum the two regression models. 

\subsection{Transformation} \label{subsec:Transform}

Let $r(c \mid x)$ denote the conditional density of $C$ given $X$ with respect to a fixed reference measure $\mu$, which in most applications will be the Lebesgue measure $\lambda$. For identifiability, it is standard to impose the assumption that the observed data satisfy \textit{coarsening at random} (CAR) with respect to the full data. CAR extends the concept of missing at random to situations where a many-to-one function of the complete data is observed. For right-censored counting processes, CAR is closely related to independent censoring defined as in~\citet{Andersen.etal:1993}, see~\citet{Gill.etal:1997} or Lemma 1 in~\citet{Munch.etal:2023}.  Here, the CAR formulation from~\citet{Van:2004} is used.
\begin{assumption} (Coarsening at random.) \label{assumption:CAR} \\
    There is a measurable function $\Tilde{r}: [0,\infty) \times \mathcal{X} \mapsto [0,\infty)$ such that
    $$r(c \mid x) = \Tilde{r}(c,x^c)$$
    for $x^c = (x(c \wedge t))_{t \geq 0}$ with $x \in \mathcal{X}$.
    \demoas
\end{assumption}
\noindent The CAR assumption is a relaxation of the stricter assumption of (conditionally) random censoring and is plausible in many real-world scenarios. For example, in a clinical trial, a patient might be more likely to drop out if their health condition worsens, where the health condition is part of the observed history $X$. This is a case of CAR, but not random censoring. 

Assume that there is a positive probability of observing the full data for any realization of $X$. This is a standard assumption for nonparametric estimation with censored data.
\begin{assumption} (Positivity.) \label{assumption:positivity} \\
It holds that $\mathbb{P}(C \geq \eta \mid X) \geq \epsilon > 0$ for some deterministic $\epsilon$.
\demoas
\end{assumption}
\noindent In Section B of the supplementary material, it is shown that the efficient influence function of $\mathbb{E}[Y(X)]$ in terms of the observed data $(C,X^C)$ is
$$\textnormal{IF} = \frac{Y(X) 1_{(C \geq \eta)}}{\mathbb{P}(C \geq \eta \mid X)} + \int_{[0,\eta)} \frac{\mathbb{E}[Y(X) \mid X^u]}{\mathbb{P}(C > u \mid X)} \left(\diff 1_{(C \leq u)} - 1_{(C \geq u)} \frac{\mathbb{P}(C \in \diff u \mid X)}{\mathbb{P}(C \geq u \mid X)} \right)-\mathbb{E}[Y(X)].$$
Observe that all quantities appearing in IF can be written as functions of $(C,X^C)$ due to CAR. Note that $\mathbb{E}[Y(X)]$ is the population version of the estimand of interest $\mathbb{E}[Y(X) \mid W]$. This motivates the DRCUT in Theorem~\ref{thm:DRCUT}. Define $\gamma(u \mid X)=r(u \mid X)/\mathbb{P}(C \geq u \mid X)$ which is the Radon-Nykodym derivative of the hazard measure for $C \mid X$ with respect to $\mu$ under $\mathbb{P}$, and define $\gamma_1$ similarly but under a different measure $\mathbb{P}_1$. 

\begin{theorem} (Doubly robust censoring unbiased transformation.) \label{thm:DRCUT} \\
    Let $\mathbb{P}_1$ and $\mathbb{P}_2$ be two probability measures, which may be thought of as candidate measures for $\mathbb{P}$. Let 
    $$Y^\ast_{\mathbb{P}_1,\mathbb{P}_2}(C,X^C) = \frac{Y(X) 1_{(C \geq \eta)}}{\mathbb{P}_1(C \geq \eta \mid X)} + \int_{[0,\eta)} \hspace{-0.1cm} \frac{\mathbb{E}_2[Y(X) \mid X^u]}{\mathbb{P}_1(C > u \mid X)} \left(\diff 1_{(C \leq u)} - 1_{(C \geq u)} \frac{\mathbb{P}_1(C \in \diff u \mid X)}{\mathbb{P}_1(C \geq u \mid X)} \right).$$
    When $\mathbb{P}$ and $\mathbb{P}_1$ satisfy Assumption~\ref{assumption:CAR} and~\ref{assumption:positivity} it holds that
    \begin{align*}
        &\mathbb{E}[Y^\ast_{\mathbb{P}_1,\mathbb{P}_2}(C,X^C) - Y(X) \mid W] \\
        &\hspace{-0.1cm} = \mathbb{E}\left[ \int_{[0,\eta)} \hspace{-0.4cm} \left(\mathbb{E}[Y(X) \mid X^u] - \mathbb{E}_2[Y(X) \mid X^u]\right) \hspace{-0.1cm}(\gamma_1(u \mid X)-\gamma(u \mid X)) \frac{\mathbb{P}(C \geq u \mid X)}{\mathbb{P}_1(C > u \mid X)} \diff \mu(u)  \mid W \right]
    \end{align*}
    and that $\Var[Y^\ast_{\mathbb{P},\mathbb{P}}(C,X^C)\mid W] \geq \Var[Y(X) \mid W]$ with equality only in the degenerate case where $\Var[Y \mid X^u]=0$ almost surely for all $u \in [0,\eta)$ where $\mathbb{P}( C \in \diff u \mid X)$ is non-zero.
\end{theorem}
\noindent The proof of Theorem~\ref{thm:DRCUT} is deferred to Section E of the supplementary material. The notation $\mathbb{E}_2$ denotes expectation under $\mathbb{P}_2$. The outcome $Y(X)$ is observed on $(C \geq \eta)$ since $Y(X)=\Tilde{Y}(X^\eta)$, so the first part of the transformation only uses complete case data but corrects for the incurred bias by reweighing with the \textit{inverse probability of censoring weights} (IPCW). This is in itself a censoring unbiased transformation and can hence be used to construct so-called IPCW pseudo-outcomes. The second term includes the contributions for the partial observations, thus making more efficient use of the data. Note that the expression in Theorem~\ref{thm:DRCUT} is not immediately well-defined due to the uncountably many null
sets associated with the conditional expectations $\mathbb{E}_2[Y(X) \mid X^u]$, and we hence take fixed regular conditional expectation throughout as in~\citet{Van:2004}. 

The transformation in Theorem~\ref{thm:DRCUT} is a generalization of the ones found in~\citet{Rubin:Van:2007} and~\citet{Steingrimsson.etal:2019}, which only apply to survival settings $X(t)=(W,1_{(T \leq t)})$ and where $C$ and
 the survival time $T$ are additionally assumed to be continuously distributed. That it simplifies to the known transformation in the survival setting may be seen by noting that both $\mathbb{P}_1(C \in \diff u \mid X)$ and $\diff 1_{(C \leq u)}$ are zero on $(T \leq u)$. The variance result appears to be new and shows that pseudo-outcomes have increased variance even if the true nuisance parameters are used.

In Section~\ref{sec:Asymp}, it is seen that $\mathbb{E}[Y^\ast_{\mathbb{P}_1,\mathbb{P}_2}(C,X^C)-Y^\ast_{\mathbb{P},\mathbb{P}}(C,X^C) \mid W]$, which is the conditional bias of the pseudo-outcomes, is important for determining the asymptotic behavior of DRCUT-based regression estimators. Theorem~\ref{thm:DRCUT} immediately implies a double robustness property as stated in Corollary~\ref{cor:DoubleRobust}, which in turn implies that Theorem~\ref{thm:DRCUT} gives a doubly robust representation for the conditional bias. The usefulness of Theorem~\ref{thm:DRCUT} comes from these observations.

\begin{corollary} (Double robustness.) \label{cor:DoubleRobust} \\
    Under the same assumptions as in Theorem~\ref{thm:DRCUT}, $\mathbb{P}_1=\mathbb{P}$ or $\mathbb{P}_2=\mathbb{P}$ implies that
    $$\mathbb{E}[Y^\ast_{\mathbb{P}_1,\mathbb{P}_2}(C,X^C) \mid W] = \mathbb{E}[Y(X) \mid W].$$
\end{corollary}

\noindent Corollary~\ref{cor:DoubleRobust} follows immediately from the first result in Theorem~\ref{thm:DRCUT}. Because the integrand relies on the product of the outcome expectation error and the censoring hazard error, correctly specifying either ensures that one of these terms is zero. Consequently, the innermost integral evaluates to zero, yielding an unbiased pseudo-outcome. In the case where $\mathbb{P}_1=\mathbb{P}_2=\mathbb{P}$, the pseudo-outcomes are referred to as \textit{oracle pseudo-outcomes}. Note that in this paper, the oracle knows the correct transformation but not the uncensored data. 

In Section A of the supplementary material, we discuss how the DRCUT approach relaxes key assumptions compared to (infinitesimal) jackknife pseudo-outcomes, the link to doubly robust random forests, and extensions of Theorem~\ref{thm:DRCUT} to estimands with dynamical conditioning information and causal estimands.

\section{Asymptotics and inference} \label{sec:Asymp}


This section focuses on the pointwise estimation of $\mathbb{E}[Y(X) \mid W=w]$ for any given $w$. This is the relevant estimand in the data application of Section~\ref{sec:Applic} since RDDs utilize conditional expectations at a specific boundary value to estimate a local causal effect. We consider estimation and large sample properties of the proposed DRCUT. A sample-splitting approach similar to~\citet{Kennedy:2023} is used since this allows one to exploit the product structure from Theorem~\ref{thm:DRCUT} to prove a rate double robustness property, permitting fast convergence rates for the estimand of interest even in settings where estimating the nuisance parameters is hard such that their individual convergence rates are slower. An example could be high dimensional settings, where regularization is used to keep the variance of the estimator from blowing up, which however makes the bias of the estimator decrease slower than it otherwise would have. In addition, sample-splitting removes the need for Donsker conditions, see~\citet{Chernozhukov.etal:2018} for additional discussions. 

Sample splitting also makes the proofs simple and model agnostic, allowing for flexible nuisance estimators whose exact statistical properties may be difficult to determine, e.g., estimators that depend on hyperparameters selected in a data-adaptive way.  Finally, when using sample-splitting, many established asymptotic results for the second-step regression method can be immediately lifted to asymptotic results about the proposed two-step estimators. This in turn means that one can leverage existing software packages for estimation and inference since the two-step estimator then asymptotically behaves as the second-step regression but where the pseudo-outcomes enter as if it was unmodified observed data, see Proposition~\ref{prop:asymp} below. Sample-splitting however has the downside that only a subset of the data is used for estimating the estimand of interest, but full-sample efficiency can be regained using cross-fitting as shown in Section~\ref{subsec:CrossFit}. 

The proposed estimation algorithm is analogous to Algorithm 1 in~\citet{Kennedy:2023} and is described in Algorithm~\ref{alg:DR}. Assume that the available data $D^{2n}$ consists of $2n$ i.i.d.\ observations. Randomly partition the data into $D^{n}_{I}$ and $D^{n}_{II}$ of size $n$ each.  Denote by $(C,X^C)$ a generic outcome drawn independently of $D^{n}_{I}$. 

\begin{algorithm}[ht!]
\caption{Pseudo-algorithm for doubly robust learning with censored data.}\label{alg:DR}
\hspace*{\algorithmicindent} \textbf{Input:} Data $D^{2n}$ split into $D^{n}_{I}$ and $D^n_{II}$.
\begin{algorithmic}[1]
\State \textbf{Nuisance estimation:} Construct estimators $\hat{\mathbb{P}}_{1,n}$ and $\hat{\mathbb{P}}_{2,n}$ of $\mathbb{P}$ using $D^{n}_{I}$.
\State \textbf{Pseudo-outcome regression:} In the sample $D^{n}_{II}$, construct the pseudo-outcomes
    \begin{align*}
        \hat{Y}^\ast_{\hat{\mathbb{P}}_{1,n},\hat{\mathbb{P}}_{2,n}}(C,X^{C}) &= \frac{Y(X) 1_{(C \geq \eta)}}{\hat{\mathbb{P}}_{1,n}(C \geq \eta \mid X)} \\
        & \quad + \int_{[0,\eta)} \frac{\hat{\mathbb{E}}_{2,n}[Y(X) \mid X^u]}{\hat{\mathbb{P}}_{1,n}(C > u \mid X^u)} \left(\diff 1_{(C \leq u)} - 1_{(C \geq u)} \frac{\hat{\mathbb{P}}_{1,n}(C \in \diff u \mid X^u)}{\hat{\mathbb{P}}_{1,n}(C \geq u \mid X^u)} \right)
    \end{align*}
    and regress them on covariates $W$, which results in a regression function
    $$\hat{\mathbb{E}}_n[\hat{Y}^\ast_{\hat{\mathbb{P}}_{1,n},\hat{\mathbb{P}}_{2,n}}(C,X^{C}) \mid D^{n}_{I}, W=w].$$
\State \textbf{Cross-fitting (optional):} Repeat steps 1 and 2, swapping the roles of $D^{n}_{I}$ and $D^{n}_{II}$. Average over the results as a the final estimate. $K$-fold cross-fitting is also possible. In the $K$-fold cross fitting case, where the data is split into $K$ equally sized folds, the number of folds used for nuisance estimation can be chosen as any fixed integer between $1$ and $K-1$, with the remaining folds being used for pseudo-outcome regression.
\end{algorithmic}
\hspace*{\algorithmicindent} \textbf{Output:} Estimator of $\mathbb{E}[Y(X) \mid W=w]$.
\end{algorithm}

\vspace{1cm}

For the sample splitting estimator, the formulas become simpler when the sample size is represented as $2n$ rather than $n$. To return to the usual setting of $n$ observations, which is for example the setting of Section~\ref{subsec:CrossFit}, one may simply replace $n$ with $n/2$ here and throughout Section~\ref{subsec:sampleSplit}.

\subsection{Sample splitting estimator} \label{subsec:sampleSplit}
For notational convenience, attention is initially restricted to the sample splitting estimator, consisting of steps 1 and 2 from Algorithm~\ref{alg:DR}. The extension to cross-fitting is given in Section~\ref{subsec:CrossFit}. We adapt the DR-learner results from~\citet{Kennedy:2023} to the censored data setting. In order to use these results, some extra notation is introduced. Introduce the shorthand $\hat{Y}^\ast = Y^\ast_{\hat{\mathbb{P}}_{1,n},\hat{\mathbb{P}}_{2,n}}$ and write $Y^\ast = Y^\ast_{\mathbb{P},\mathbb{P}}$ for the oracle pseudo-outcomes. Introduce the conditional bias of the pseudo-outcomes
\begin{align*}
    \hat{b}(w;D^{n}_{I}) &= \mathbb{E}[\hat{Y}^\ast(C,X^{C}) - Y^\ast(C,X^{C}) \mid D^{n}_{I}, W=w ].
\end{align*}
The effect of conditioning on $D^{n}_{I}$ is that $\hat{\mathbb{P}}_{1,n}$ and $\hat{\mathbb{P}}_{2,n}$ are fixed in the conditional expectation. Define:
\begin{align*}
    m(w) &= \mathbb{E}[Y^\ast(C,X^C) \mid W=w], \\
    \hat{m}(w) &=\hat{\mathbb{E}}_n[ \hat{Y}^\ast(C,X^{C}) \mid D^{n}_{I}, W=w], \\
    \tilde{m}(w) &= \hat{\mathbb{E}}_n[Y^\ast(C,X^C) \mid W=w].
\end{align*}
Thus, $m(w)$ is the oracle conditional expectation which also equals $\mathbb{E}[Y(X) \mid W=w]$, $\hat{m}(w)$ is the regression estimator obtained from regressing $\hat{Y}^\ast(C,X^{C})$ on $W$ in the sample $D^{n}_{II}$ using a given regression estimator $\hat{\mathbb{E}}_n$, and $\tilde{m}(w)$ is the oracle regression estimator. To infer the asymptotics of $\hat{m}(w)$, decompose $\hat{m}(w) - m(w)$ into the sum of $\hat{m}(w) - \Tilde{m}(w)$ and $\Tilde{m}(w) - m(w)$.
The asymptotics of $\Tilde{m}(w) - m(w)$ can often be inferred from known asymptotic theory for the chosen regression estimator $\hat{\mathbb{E}}_n$. From hereon, it is assumed that the convergence rate is $n^{-\alpha}$.
\begin{assumption} (Convergence rate of second-step regression method.) \label{assumption:convRate}\\
    It holds that $\Tilde{m}(w) - m(w) = O_{\mathbb{P}}(n^{-\alpha})$.
    \demoas
\end{assumption}

\noindent For the remaining terms, write
\begin{align*}
    \hat{m}(w)-\Tilde{m}(w) &= \hat{\mathbb{E}}_n[\hat{b}(W;D^{n}_{I}) \mid D^{n}_{I}, W=w] \\
    &+ \left(\hat{m}(w)-\Tilde{m}(w)-\hat{\mathbb{E}}_n[\hat{b}(W;D^{n}_{I}) \mid D^{n}_{I}, W=w]\right).
\end{align*}
Theorem~\ref{thm:DRCUT} gives a doubly robust representation for the conditional bias $\hat{b}(W;D^{n}_{I})$. Consequently, as we show later, the first term is usually $o_\mathbb{P}(n^{-\alpha})$ if at least one of the nuisance estimators is consistent and converges sufficiently fast. It is this property that allows us to establish rate double robustness of the approach. Following Definition 1 of~\citet{Kennedy:2023}, the regression method $\hat{\mathbb{E}}_n$ is said to be \textit{stable} if the second term is $o_\mathbb{P}(n^{-\alpha})$ whenever $d(\hat{Y}^\ast,Y^\ast) = o_\mathbb{P}(1)$ for a suitable stochastic distance $d$. 

Let $(C_i,X_{i}^{C_i})$ be the $i$'th observation in $D^{n}_{II}$. By Theorem 1 in~\citet{Kennedy:2023}, the class of linear smoothers
$$\hat{\mathbb{E}}_n[ f(C,X^{C}; D^{n}_{I}) \mid D^{n}_{I}, W=w] = \sum_{i=1}^n p_i(w;W^n) f(C_{i},X^{C_{i}}_{i}; D^{n}_{I})$$
for $W^n = (W_k)_{1 \leq k \leq n}$ is stable under suitable regularity conditions. Some prominent methods that belong to this class are listed in~\citet{Kennedy:2023}. Importantly, local linear regression is a linear smoother, which is the de facto method used in RDDs and which is also employed in Section~\ref{sec:Sim} and~\ref{sec:Applic}. It is also possible to force the output of more flexible methods to be on this form to keep inference tractable and enhance interpretability. This can for example be done following~\citet{Verdinelli:Wasserman:2021} which also relies on sample splitting, first fitting a random forest and then using the resulting estimator to define a kernel used for local linear regression in the second split. While any regression method can be used to regress the pseudo-outcomes on covariates in Algorithm~\ref{alg:DR}, we only prove asymptotic results for linear smoothers.

To define the distance under which linear smoothers are stable, first introduce the conditional $L^2(\mathbb{P})$-norm 
$$\lVert f(Z) \rVert_{w,D^{n}_{I}} = \mathbb{E}[f(Z)^2 \mid D^{n}_{I}, W=w]^{1/2}.$$
Theorem 1 in~\citet{Kennedy:2023} then implies that linear smoothers are stable at $W=w$ with respect to the stochastic distance $d_{w,D^{2n}}$ given by
$$d_{w,D^{2n}}(g,f) = \sum_{i=1}^n \left\{ \frac{p_i(w;W^n)^2}{\sum_{j=1}^n p_j(w;W^n)^2} \lVert g(C,X^{C};D^{n}_{I}) - f(C,X^{C}) \rVert_{W_{i},D^{n}_{I}}^2  \right\}$$
whenever $d_{w,D^{2n}}(0,\Var[Y^\ast(C,X^C) \mid W= \bigcdot \:])^{-1} = O_{\mathbb{P}}(1)$. Thus, if $d_{w,D^{2n}}(\hat{Y}^\ast,Y^\ast) = o_{\mathbb{P}}(1)$ and $d_{w,D^{2n}}(0,\Var[Y^\ast(C,X^C) \mid W= \bigcdot \:])^{-1} = O_{\mathbb{P}}(1)$ then stability gives
$$\hat{m}(w)-\Tilde{m}(w) = \hat{\mathbb{E}}_n[\hat{b}(W;D^{n}_{I}) \mid D^{n}_{I}, W=w] + o_\mathbb{P}(n^{-\alpha}).$$
One can thus focus on the asymptotics of the conditional bias. Due to its product structure, one can obtain rate double robustness results like the one in Proposition~\ref{prop:asymp}. Introduce the stochastic norm 
$$\lVert f(u,X;D^{n}_{I}) \rVert_{2,w,D^{2n}} = \left\{ \sum_{i=1}^n \frac{\vert p_i(w;W^n) \vert}{\sum_{j=1}^n \vert p_j(w;W^n) \vert}  \int_{[0,\eta)} \lVert f(u,X;D^{n}_{I}) \rVert_{W_i,D^{n}_{I}}^2 \diff \mu(u)  \right\}^{1/2}.$$

\begin{proposition} (Rate double robustness under weighted $L^2$-rates.) \label{prop:asymp} \\
    Impose the assumptions from Theorem~\ref{thm:DRCUT}, Assumption~\ref{assumption:convRate}, and for a fixed $w$
    \begin{enumerate}[label=(\roman*)]
        \item $\inf_z \Var[Y(X) \mid W=z] > 0$;
        \item $d_{w,D^{2n}}(\hat{Y}^\ast,Y^\ast) = o_{\mathbb{P}}(1)$;
        \item $\sum_{i=1}^n \vert p_i(w;W^n) \vert = O_{\mathbb{P}}(1)$;
        \item $\lVert \mathbb{E}[Y(X) \mid X^u] - \hat{\mathbb{E}}_{2,n}[Y(X) \mid X^u]  \rVert_{2,w,D^{2n}} = O_{\mathbb{P}}(n^{-\alpha_1})$;
        \item $\lVert \hat{\gamma}_{1,n}(u \mid X)-\gamma(u \mid X)  \rVert_{2,w,D^{2n}} = O_{\mathbb{P}}(n^{-\alpha_2})$;
        \item $\alpha_1 + \alpha_2 > \alpha$.
    \end{enumerate}
     Then $\hat{m}(w)-m(w) = \tilde{m}(w)-m(w) + o_{\mathbb{P}}(n^{-\alpha})$ i.e., oracle efficiency is obtained.
\end{proposition}
\noindent The proof of Proposition~\ref{prop:asymp} is deferred to Section F of the supplementary material. Condition (i) is a non-degeneracy assumption, and is used to prove $d_{w,D^{2n}}(0,\Var[Y^\ast(C,X^C) \mid W= \bigcdot \:])^{-1} = O_{\mathbb{P}}(1)$. Together with (ii), this implies stability of linear smoothers. Condition (ii) is a consistency condition for the pseudo-outcomes, which can be expected to hold when condition (iv) and (v) are imposed for positive $\alpha_1$ and $\alpha_2$. As noted in~\citet{Kennedy:2023}, many linear smoothers satisfy that $\sum_{i=1}^n \vert p_i(w ; W^n) \vert$ is bounded by a fixed constant with probability one, which would imply (iii). Condition (iv) and (v) are assumptions about the weighted $L^2(\mathbb{P})$ convergence rates of the outcome and censoring estimators, respectively. To obtain reasonable convergence rates in (iv) and (v), one likely needs to assume that the dependence on $X^u$ can be captured by a $d$-dimensional stochastic process $Z(u) = f(u,X^u)$. Assuming that the $L^q(\mathbb{P})$ convergence rates results from~\citet{Stone:1980,Stone:1982} carry over to the weighted $L^2(\mathbb{P})$-norm $\lVert \cdot \rVert_{2,w,D^{2n}}$ and that $z \mapsto \mathbb{E}[Y(X) \mid Z(u)=z]$ is $s$-times continuously differentiable, Theorem 1 in~\citet{Stone:1980,Stone:1982} implies that the optimal convergence rate in a minimax sense is $O_{\mathbb{P}}(n^{-r} )$ for $r=1/(2+d/s)$ under some regularity conditions. This rate can be obtained using, e.g., series or local polynomial estimators. Other structured assumptions, such as sparsity or additivity, are popular alternatives to smoothness when they are applicable, see for instance~\citet{Yang:Tokdar:2015}.

Proposition~\ref{prop:asymp} says that the approach achieves rate double robustness. Furthermore, if (ii) can be satisfied with either $\alpha_1=0$ or $\alpha_2=0$, then model double robustness is also achieved, so one inconsistent estimator is allowed if the other converges sufficiently quickly. The results of the simulation study in Section~\ref{sec:Sim} indicate that model double robustness is achieved in that setting. As discussed in Remark 5 of~\citet{Kennedy:2023}, results like Proposition~\ref{prop:asymp} are important for inference since they imply that the asymptotic distribution when using estimated and oracle pseudo-outcomes are identical. For example, if $\tilde{m}(w)$ is asymptotically Gaussian then so is $\hat{m}(w)$ and their asymptotic mean and variance are the same. Confidence intervals for $m(w)$ may hence be constructed by treating the estimated pseudo-outcomes as if they were observed outcomes and employing the usual asymptotic distributional approximation. Standard implementations can therefore be used. 

\begin{remark} (Extension to vector-valued outcomes.) \label{rmk:VectorValued} \\
   The results and proofs of Section~\ref{subsec:Transform} are unchanged if $Y$ takes values in $\mathbb{R}^p$ for some $p \geq 1$ rather than $p=1$. Similarly, if Proposition~\ref{prop:asymp} holds for each coordinate of $Y$, then also $\hat{m}(w)-m(w)=\tilde{m}(w)-m(w)+o_{\mathbb{P}}(n^{-\alpha})$ as random vectors so in this case the joint asymptotic distribution are the same by Slutsky's lemma. \demormk
\end{remark}

\noindent Under smoothness assumptions on $\mathbb{E}[Y(X) \mid W=w]$, local polynomial regression has the minimax optimal convergence rate, meaning that the optimal rate is also obtained using Algorithm~\ref{alg:DR} when oracle efficiency is obtained and local polynomial regression is used as the second-step estimator. Note however that the sample size used in the asymptotic approximation becomes $n$ rather than $2n$ which was the number of observations that was originally available, leading to a suboptimal constant in the minimax risk of the estimator. As shown in the next section, cross-fitting may be used to regain full sample efficiency. It is left to future work to determine how far the resulting constant is from the optimal, see~\citet{Fan:1993} for similar considerations in the case of local linear regression with no censoring.

\subsection{Cross-fitted estimator} \label{subsec:CrossFit}
In this section, the arguments are extended to $K$-fold cross-fitting. Assume that one has access to $n$ observations and deterministically partition the data into folds of size $n/K$. For simplicity, assume that $\hat{\mathbb{E}}_n$ is asymptotically Gaussian such that $n^{\alpha}\{\tilde{m}(w)-m(w)\} \rightarrow \mathcal{N}(\mu,\sigma^2)$. Let $\tilde{m}_k(w)$ be the oracle estimator when only data from fold $k$ $(k=1,\dots,K)$ is used. Let $D^{-k}$ be the data not in fold $k$. Write $\hat{Y}^\ast_{-k}$ for the pseudo-outcomes with estimates based on $D^{-k}$. Similarly, $\hat{\mathbb{E}}_{k}$ is the regression estimator based on the data in fold $k$.  Denote by
$$\hat{m}_k(w) = \hat{\mathbb{E}}_{k}[\hat{Y}^\ast_{-k}(C,X^C) \mid D^{-k}, W=w]$$
the estimator obtained from estimating the nuisance parameters using $D^{-k}$ and then regressing over the pseudo-outcomes from fold $k$. The proposed cross-fitted estimator is then $\hat{m}^{\textnormal{CF}}(w) =  1/K\sum_{k=1}^K \hat{m}_k(w)$. This cross-fitting scheme is similar to DML1 in~\citet{Chernozhukov.etal:2018}. An alternative not explored here could be like DML2 to first compute all the pseudo-outcomes $\hat{Y}^\ast_{-k}(C,X^C)$ and then input them simultaneously into $\hat{\mathbb{E}}_n$. The following proposition shows that the cross-fitted estimator regains full-sample efficiency.

\begin{proposition}(Asymptotic distribution of cross-fitting estimator.) \label{prop:crossFit} \\
    Under the assumptions from Proposition~\ref{prop:asymp} it holds that 
    \begin{align*}
    n^{\alpha}\{\hat{m}^{\textnormal{CF}}(w) - m(w)\} \rightarrow \mathcal{N}(K^\alpha \mu,K^{2\alpha-1} \sigma^2)
\end{align*}
in distribution.
\end{proposition}

\noindent The proof of Proposition~\ref{prop:crossFit} is deferred to Section G of the supplementary material. The maximal convergence rate is usually $\sqrt{n}$, so we can assume $\alpha \leq 1/2$. Consequently, the asymptotic variance of the cross-fitted estimator is no larger than that of the full-sample oracle estimator $\tilde{m}(w)$ and is strictly less when $\alpha < 1/2$, while the bias is increased by a factor of $K^\alpha$. In the special case where $\mu=0$ and $\alpha=1/2$, the asymptotic distribution of $\tilde{m}(w)$ and $\hat{m}^{\textnormal{CF}}(w)$ are identical which agrees with previous results in the literature, see e.g., Remark 3.1 and Theorem 3.1 in~\citet{Chernozhukov.etal:2018}. Proposition~\ref{prop:crossFit} implies that the standard error of $\hat{m}^{\textnormal{CF}}(w)$ can be estimated by averaging the estimated standard errors $\hat{\sigma}_k / (n/K)^\alpha$ of $\hat{m}_k(w)$ ($k=1,\dots,K$) and scaling by $K^{-1/2}$.

In Section A of the supplementary material, we discuss implied convergence rates for the MSE, results on oracle efficiency without sample splitting, and bias-variance tradeoffs in cross-fitting.

\section{Simulations} \label{sec:Sim}

\subsection{Data-generating process} \label{subsec:SimGen}

\begin{figure}[b]
\centering
\scalebox{0.75}{
   \begin{tikzpicture}[node distance=8em, auto]
	\node[punkt] (g) {Healthy};
	\node[right=4cm of g] (i1) {};
        \node[punkt, above=0.5cm of i1] (i2) {Illness};
        \node[punkt, below=0.5cm of i1] (i3) {Dead};
        \node[anchor=north east, at=(g.north east)] {$1$};
        \node[anchor=north east, at=(i2.north east)] {$2$};
        \node[anchor=north east, at=(i3.north east)] {$3$};
	\path (g) edge [pil] node [above=0.25cm]  {$\mu_{12}$} (i2)
	;
        \path (g) edge [pil] node [below=0.25cm]  {$\mu_{13}$} (i3)
	;
        \path (i2) edge [pil] node [right=0.25cm]  {$\mu_{23}$} (i3)
	;
    \end{tikzpicture}
}
\caption{The irreversible illness-death model for the process $Z$. Transitions from state $j$ to state $k$ has the transition hazard $\mu_{jk}$.}
\label{fig:IllnessDeath}
\end{figure}
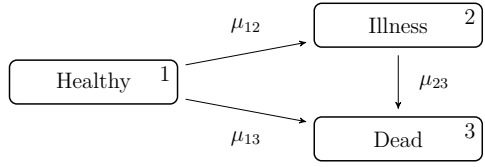

To examine the finite sample predictive and inference performance of the proposed estimator and to demonstrate the double robustness property, a numerical study is conducted. The complete-case data is specified as $X=(Z,W)$ where $Z$ follows the irreversible illness-death model depicted in Figure~\ref{fig:IllnessDeath} with a time-horizon of $\eta=5$ and initial state $Z(0)=1$ and the baseline covariate is $W \sim \textnormal{Uniform}(-4,4)$. There is state-dependent censoring with drop-out only occurring from state 1. The outcome of interest is the duration spent in the illness state before the end of the observation window meaning $Y(X)=\int_{[0,\eta)} 1\{Z(s) = 2\} \diff s$ and $\mathbb{E}[Y(X) \mid W] = \int_{[0,\eta)} p_{2}(s,W) \diff s$ for the state-occupation probability $p_{2}(s, W) = \mathbb{P}(Z(s) = 2 \mid W)$. The censored outcome $(C,X^C)$ is simulated by first simulating $W$ and then simulating $(C,Z^C) \mid W$ using Lewis’ thinning algorithm from~\citet{Ogata:1981}. 
A total of $500$ samples of sizes $n \in \{1000,5000,10000,30000\}$ are considered. The specific functional forms of the transition and censoring hazards and the numerical scheme used to calculate the DRCUT $Y^*$ are provided in Section C of the supplementary material. The \texttt{R} implementation~\citep{R:2023} is available on GitHub (\url{https://github.com/oliversandqvist/Web-appendix-drcut}). 

\begin{remark} (Relevance of the estimand.) \\
    This setup is motivated by disability insurance applications, where the length of an illness is a key driver of expenses since insureds often receive disability benefits as long as they are disabled to make up for lost wages. Since they receive large benefits, subjects do not leave the portfolio while disabled, so there is no censoring hazard while in the illness state. This could also be a relevant estimand in medical applications where the length of an illness or a hospital stay could be an important outcome.
    \demormk
\end{remark}

\subsection{Estimators}\label{subsec:SimEst}

The following estimators are considered: 
\begin{enumerate}[label=(\arabic*)]
    \item A plug-in estimator with $p_2$ estimated by the Conditional Aalen-Johansen (CAJ) of~\citet{Bladt:Furrer:2023} using the \texttt{R}-package \texttt{AalenJohansen};
    \item A plug-in estimator with transition hazards $\mu_{12}$, $\mu_{13}$, and $\mu_{23}$ estimated by the Highly Adaptive Lasso (HAL) of~\citet{Benkeser:Van:2016} and~\citet{Munch.etal:2024} using a custom implementation relying on the \texttt{R}-package \texttt{glmnet};
    \item Two-fold cross-fitted doubly robust pseudo-outcomes with  transition and censoring hazards estimated by HAL;
    \item Two-fold cross-fitted doubly robust pseudo-outcomes with  transition hazards estimated by HAL and censoring hazards estimated by a misspecified parametric model;
    \item Two-fold cross-fitted doubly robust oracle pseudo-outcomes;
    \item Estimators (3), (4), and (5) but with IPCW pseudo-outcomes. 
\end{enumerate}
The simulation study is designed such that estimator (1) is misspecified. While the transition hazards in (2)-(4) are all (approximately) correctly specified, the censoring hazard is correctly specified in (3) but misspecified in (4). Estimator (5) is an oracle method using true hazards. The IPCW variants in (6) thus serve as the correctly specified, misspecified, and oracle benchmarks. 

For all pseudo-outcome-based methods, the second-step regression method is chosen as a local linear regression using \texttt{lprobust} from the \texttt{R}-package \texttt{nprobust} with accompanying paper~\citet{Calonico.etal:2019} using default parameters except for the bandwidth. With a one-dimensional covariate, the MSE and MISE optimal bandwidths satisfy $h \propto n^{-1/5}$ and lead to reasonable finite sample performance. However, to obtain a non-vanishing bias in the asymptotic Gaussian distribution, one needs $nh^5 \rightarrow 0$ (undersmoothing) or an explicit bias correction.  For further details and discussions, see~\citet{Calonico.etal:2019} and the references therein. We proceed via undersmoothing, first using a separate simulation to find a bandwidth with good performance when $n=5\,000$ and then letting $h \propto n^{-1/4.5}$. In this case, the convergence rate of $\hat{\mathbb{E}}_n$ is $\sqrt{nh} \propto n^{7/18}$ so $\alpha=7/18$.

The custom implementation of HAL for hazard estimation is a modification of the code from~\citet{Rytgaard.etal:2022} and~\citet{Rytgaard.etal:2023} allowing for higher-order interactions than second-order which is needed for estimation of $\mu_{23}$. HAL is chosen since it is a general-purpose estimator that in~\citet{Benkeser:Van:2016} is demonstrated to have reasonable empirical performance both in smooth and discontinuous settings and is shown to have desirable asymptotic properties in~\citet{Munch.etal:2024} whenever the true function is multivariate càdlàg. With this choice of hazard rates, HAL is expected to estimate the censoring hazard very closely since the true hazard is piecewise constant, while it is expected to have a harder time estimating the transition hazards of the illness-death model since these are more complicated.    

The misspecified parametric family for the censoring hazard is chosen as the parametric family where $X \mid C$ has hazard $\gamma(t,W;\beta) = 1\{Z(t)=1\} \exp(\beta_1+\beta_2 \times t + \beta_3 \times W)$. This is expected to have poor performance since the true function $\gamma(t,W)$ is a symmetric step-function. The CAJ estimator is expected to be biased since the model is non-Markovian and the censoring is state-dependent, confer with Assumption 2 and Remark 2.1 in~\citet{Bladt:Furrer:2023}. Similarly to~\citet{Munch.etal:2023} and~\citet{Gunnes.etal:2007}, we observed that highly non-Markovian behaviour as well as high degrees of state-dependent censoring were required for the bias of the CAJ estimator to be sizeable, and it further seems that the effect of covariates has to be small compared to the non-Markovianity of $Z$ and the state-dependence of $C \mid X$.

\begin{remark} (Marginal estimands for the irreversible illness-death model.) \\
    The paper~\citet{Munch.etal:2023} also uses efficient-influence-function-based estimators of estimands formulated using multi-state models and employs HAL to estimate nuisance parameters. Their results are however specialized to marginal state-occupation probabilities in the illness state for an illness-death model. This paper can be viewed as an extension of their approach to any square-integrable outcome and, more importantly, an extension to estimands that may depend on baseline covariates. As discussed in Section A of the supplementary material, our approach can further be generalized to accommodate conditioning on the history of the process.
    \demormk
\end{remark}

\subsection{Results}\label{subsec:SimResults}

\begin{figure}[t]
\centering
\begin{minipage}{.5\linewidth}
  \centering
  \includegraphics[width=1\linewidth]{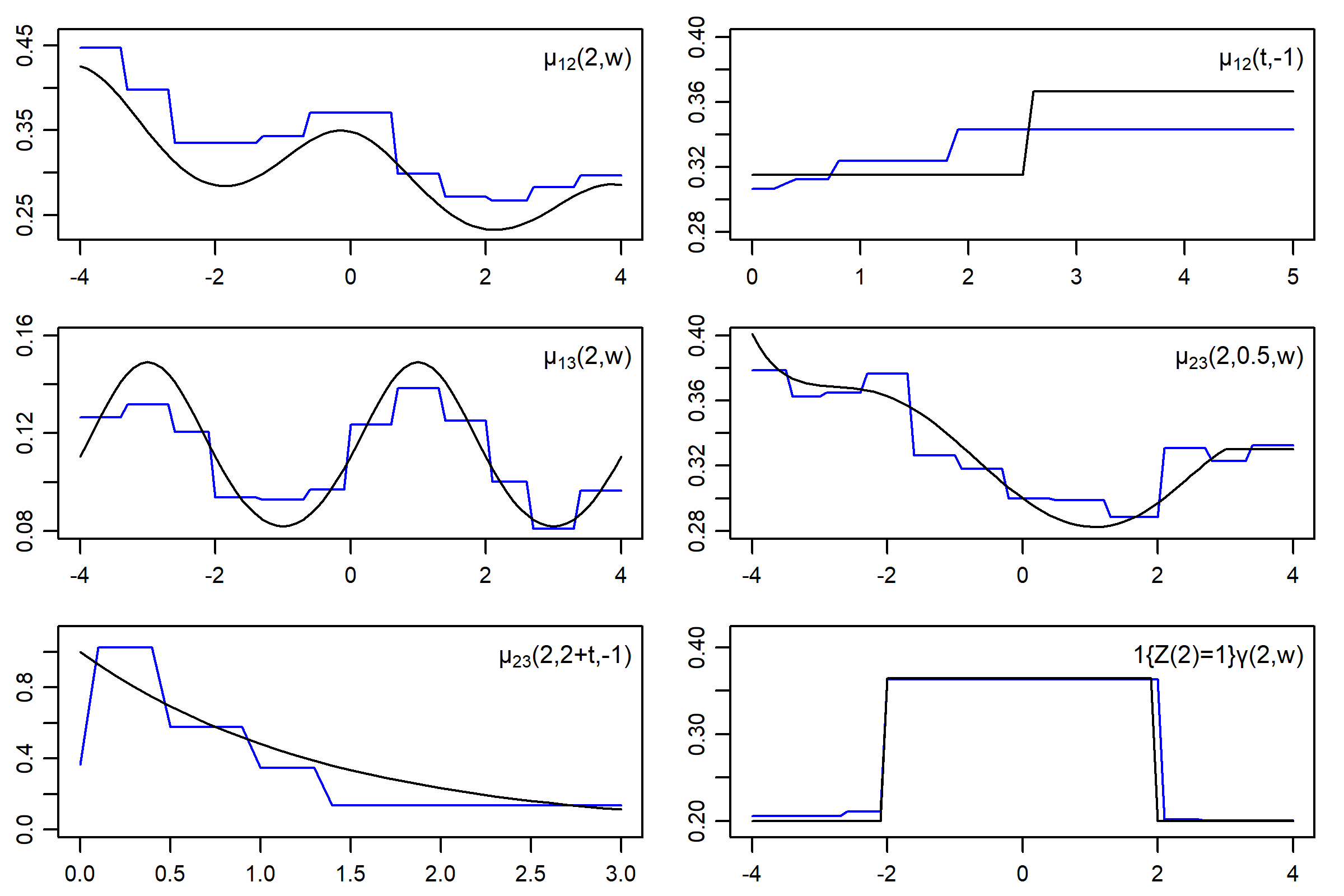}
\end{minipage}%
\begin{minipage}{.5\linewidth}
  \centering
  \includegraphics[width=1\linewidth]{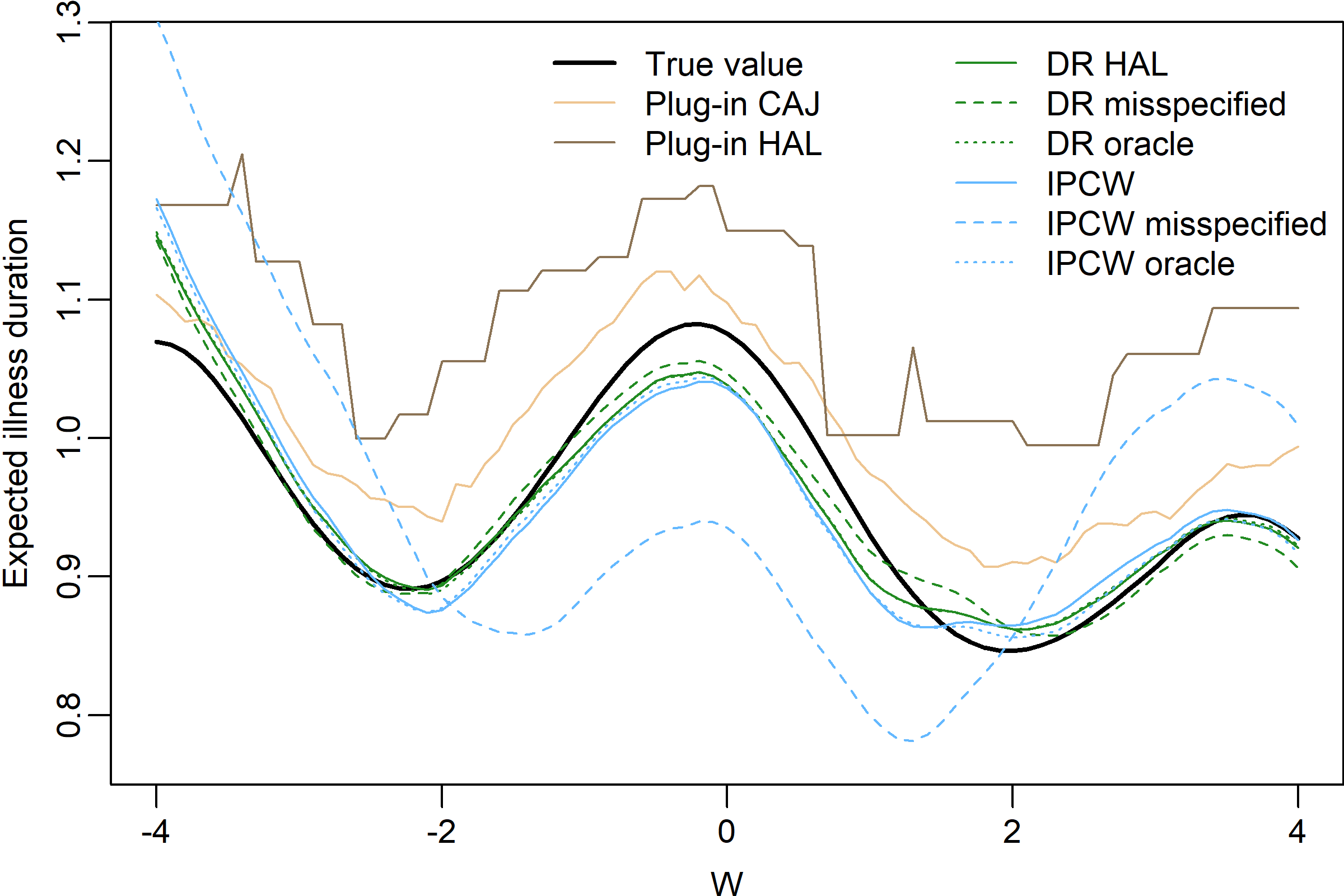}
\end{minipage}
\caption{\textbf{Left Panel:} Fitted HAL estimates and actual hazards at specific input values indicated at the top right corner for a single simulation. \textbf{Right Panel:} Estimators and true value of $\mathbb{E}[Y(X) \mid W]$ as a function of $W$ for a single simulation.}
\label{fig:EstimandSingleSim}
\end{figure}

\begin{figure}[t]
    \centering
    \includegraphics[width=0.5\linewidth]{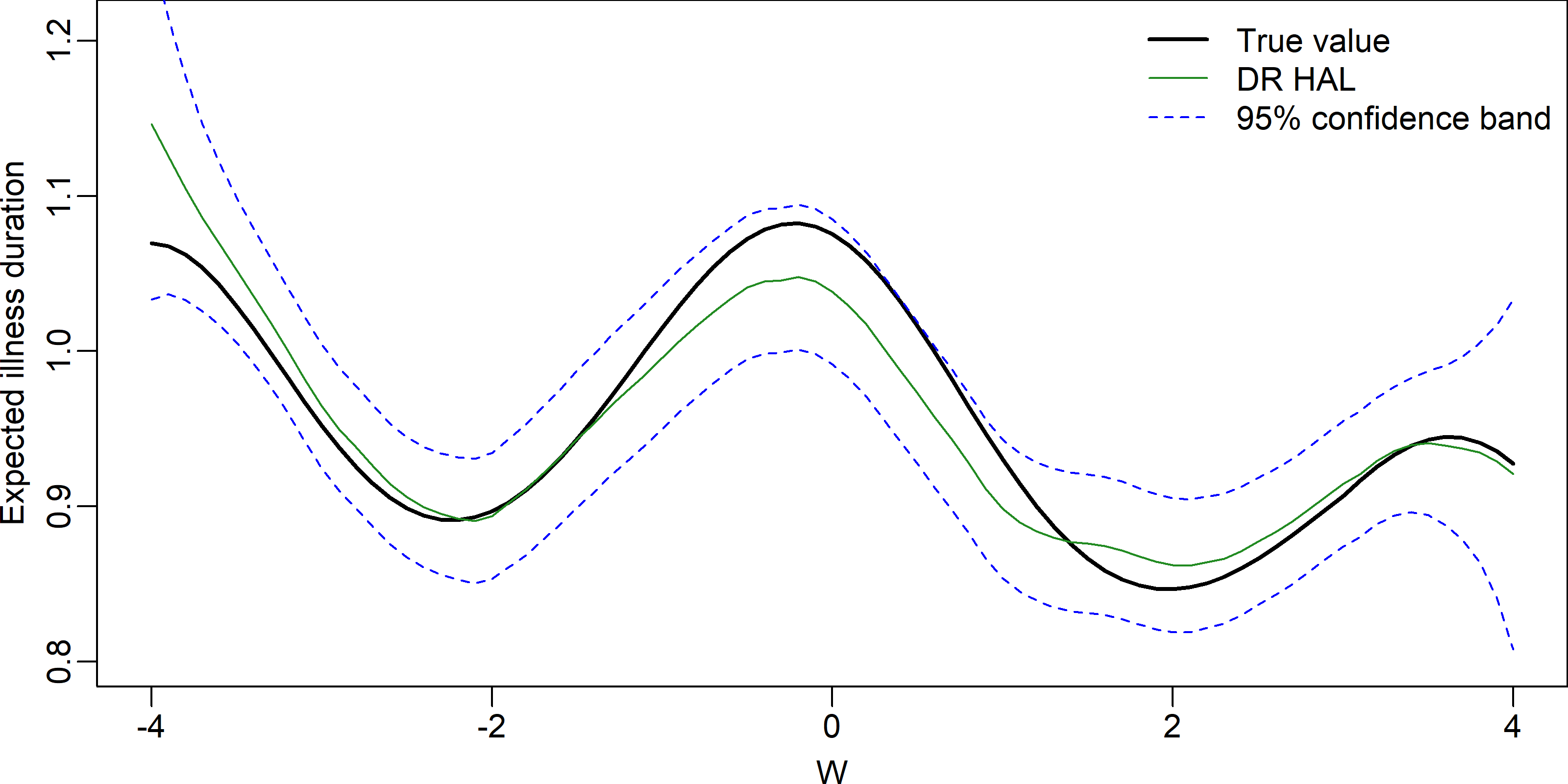}
    \caption{Estimates using two-fold cross-fitted doubly robust pseudo-outcomes with nuisance parameters estimated by HAL (green), the true curve (black), and pointwise $95\%$ confidence bands outputted by \texttt{lprobust} (blue) for a single simulation. }
    \label{fig:CISingleSim}
\end{figure}

The results for the first simulation are depicted in Figure~\ref{fig:EstimandSingleSim} and~\ref{fig:CISingleSim}. The estimand is a lower dimensional and smoother object than the individual hazards which makes it possible to nonparametrically estimate at a faster rate than the hazards. Since the pseudo-outcome methods use local linear regression as the second step estimator, this additional structure is exploited and these methods are therefore expected to perform well as long as the pseudo-outcomes are close to their oracle counterparts. 

As seen on the left part of Figure~\ref{fig:EstimandSingleSim}, HAL captures the general shape of the transition hazards reasonably well and the censoring hazard extremely well as was expected. However, the right plot shows that the plug-in estimator based on HAL-estimated transition hazards performs poorly. HAL employs regularization to balance bias and variance to be optimal for the individual hazards, but this bias-variance trade-off is seen to be suboptimal for the estimand of interest as the estimate becomes too biased. The CAJ estimator is biased in this setting, and this bias carries over to the plug-in estimator, but the estimator still performs better than the plug-in HAL estimator. The HAL-based IPCW estimator performs well and is almost indistinguishable from the oracle IPCW estimator which is unsurprising since the estimated censoring hazard is very close to the true value. As expected, the misspecified IPCW estimator performs poorly. The doubly robust pseudo-outcomes perform well, resulting in values similar to those of the non-misspecified IPCW pseudo-outcomes. For the HAL-based doubly robust pseudo-outcomes, one might have suspected that this was solely a consequence of the good performance of the IPCW term, but then the doubly robust pseudo-outcomes with a misspecified censoring hazard should have performed poorly which is not the case. For those pseudo-outcomes, one sees the remarkable phenomenon that although both the HAL-estimated transitions hazards and the misspecified censoring hazard gave poor estimates by themselves, the doubly robust property of the pseudo-outcomes makes them perform well when used jointly. For Figure~\ref{fig:CISingleSim}, one sees that the true curve is always contained in the pointwise $95\%$ confidence bands albeit barely around $W=0.5$.

\begin{figure}[t]
    \centering
\includegraphics[width=1\linewidth]{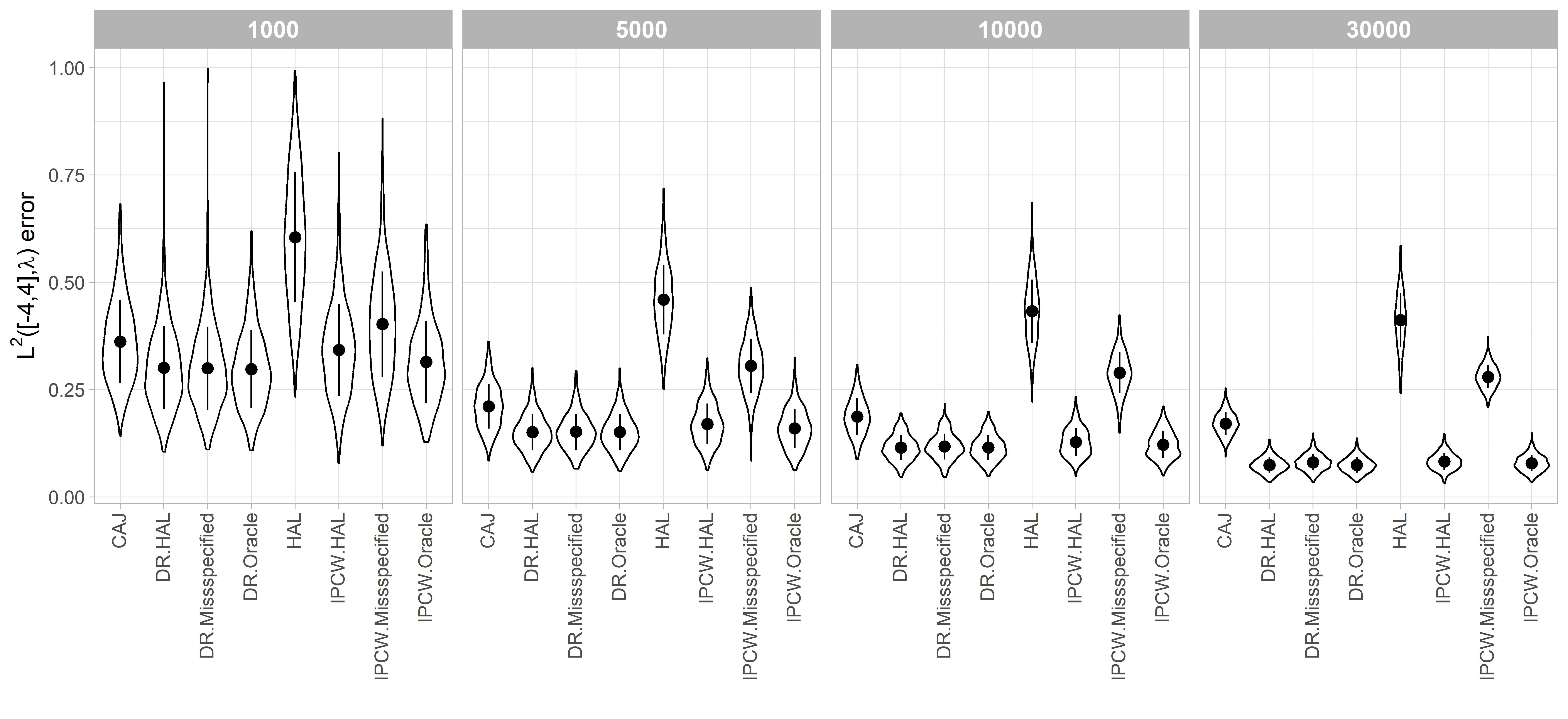}
    \caption{Violin plot of the $L^2([-4,4],\lambda)$ error for the different estimators and values of $n$ with Mean $\pm$ Standard deviation indicated as a point range using 500 simulations.}
    \label{fig:Violin}
\end{figure}

Similar patterns emerge across the $500$ simulations. The reported performance metrics are the $L^2([-4,4],\lambda)$ error which is relevant for prediction, and the empirical coverages of the confidence intervals for methods (3) and (5) which is relevant for inference. Additional performance metrics for prediction were computed, but their results were qualitatively highly similar to the $L^2([-4,4],\lambda)$ error and are hence not reported. Figure~\ref{fig:Violin} leads to many of the same qualitative conclusions as Figure~\ref{fig:EstimandSingleSim} regarding which estimators perform well and how similar they are. Additionally, one can see that the average performance of the plug-in CAJ, plug-in HAL, and misspecified IPCW estimator does not improve noticeably after $n=5\,000$ although the variability decreases. For the remaining estimators, both the average performance and variability improves as $n$ increases, and their densities are similar. 

\begin{figure}[t]
\centering
\begin{minipage}{.5\linewidth}
  \centering
  \includegraphics[width=1\linewidth]{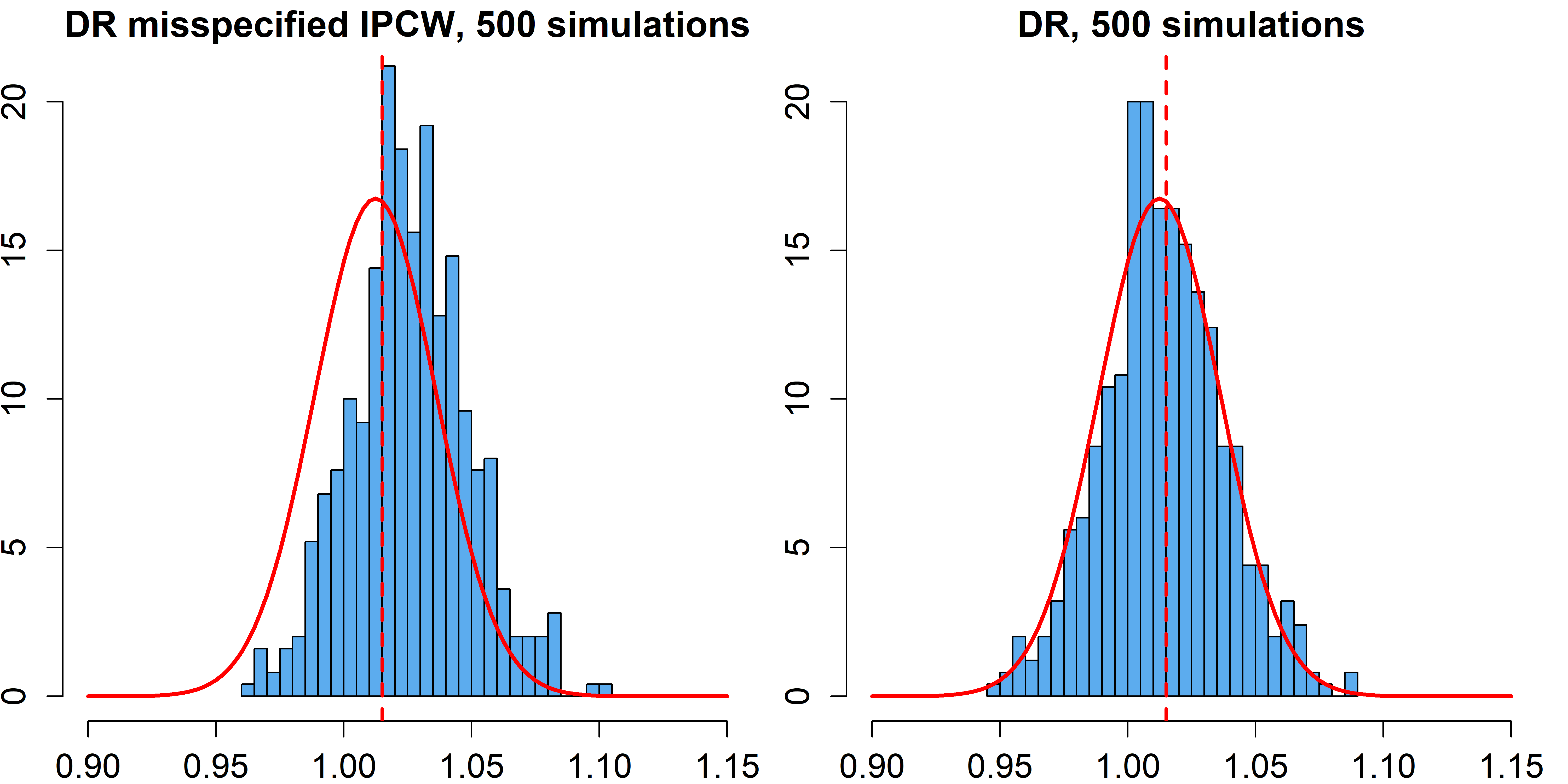}
\end{minipage}%
\begin{minipage}{.5\linewidth}
  \centering
  \includegraphics[width=1\linewidth]{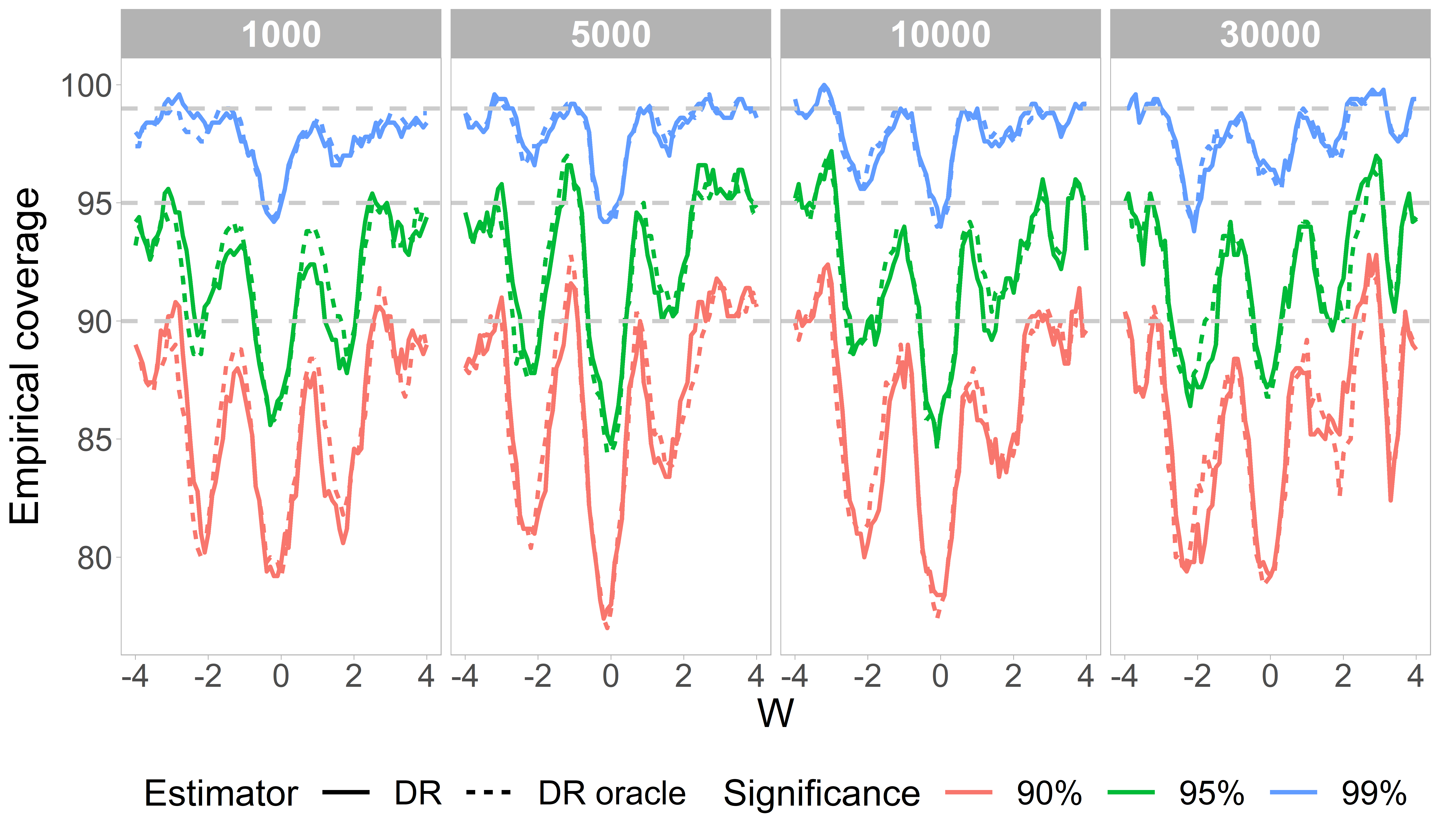}
\end{minipage}
\caption{\textbf{Left Panel:} Histogram over estimates at the point $W=-1$ for the doubly robust pseudo-outcomes where the censoring hazards are estimated with HAL and a misspecified parametric family.  The Gaussian approximation in red is obtained from the oracle pseudo-outcomes and the dashed red line is the true value of the estimand. Based on 500 simulations of size $n=30\,000$. \textbf{Right Panel:} The empirical coverages of the $99\%$ (blue), $95\%$ (green), and $90\%$ (red) confidence intervals using a Gaussian approximation with standard errors obtained from \texttt{lprobust} using HAL-estimated and oracle doubly robust pseudo-outcomes. The nominal values are highlighted with dashed grey lines.}
\label{fig:inference}
\end{figure}

Although the performance of the doubly robust pseudo-outcomes with a misspecified and HAL-estimated censoring hazard appear similar in terms of predictive performance, it can be seen from the left plot in Figure~\ref{fig:inference} that the one using HAL agrees better with the Gaussian distributional approximation obtained from the oracle values and also with the true value of the estimand. From the right plot, one can see that the empirical coverages of the confidence intervals deviate somewhat from their nominal values, but much more importantly for this study is that the confidence intervals for the oracle and estimated doubly robust pseudo-outcomes are highly similar especially for $n \geq 5\,000$ suggesting that oracle efficiency is obtained. The coverages are close to their nominal value when $W$ is away from $-2$, $0$, and $2$, where the curvature of the true estimand is the greatest, which indicates that the chosen bandwidth might have led to too much smoothing for these values of $W$. 

The choice of bandwidth greatly affects the validity of inference based on kernel estimators, see for example Table I in~\citet{Calonico.etal:2014}, making bandwidth selection important. In a setting resembling this numerical study, it would hence be natural to select different bandwidths for different values of $W$. It would be desirable to do this in some data-adaptive way, but then the resulting regression estimator might fall outside the class of linear smoothers and hence also outside of Proposition~\ref{prop:asymp}. It would be of great interest to extend the results of the present paper to such estimators, but this is left to future work.

\section{Data application} \label{sec:Applic}

The proposed method is demonstrated by an application to data from LSYPE, Waves 1 to 5 (\cite{CLS:2024},~\cite{Calderwood:Sanchez:2016}). LSYPE is a panel survey of initially around $16\,000$ young people (YP) born between September 1989 and August 1990 in England. Data was collected starting in 2004 and the first five Waves consisted of annual interviews with the YP and their carers. Thus, YP were in Year 9 during Wave 1. YP were allowed to leave school after Year 11 with post-compulsory schooling consisting of Years 12 and 13. We aim to estimate the impact of the Education Maintenance Allowance (EMA), a conditional cash transfer program, on time spent in full-time education. This is achieved by using the proposed methods to construct an RDD in the presence of censored data.

This data application is chosen to be relatively different from the setting in the simulation study in order to demonstrate the flexibility of the methodology. Firstly, YP are subject to intermittent observations while the simulation study considered data available in continuous time, but both can be accommodated in the setup since almost no assumptions have been placed on the reference measure $\mu$ and the distribution of $X$. This in turn makes it natural to use other nuisance estimators than HAL in the data application, which is again allowed due to the model-agnostic nature of our results. Secondly, the estimand of interest in the data application is a transformation of several conditional expectations, rather than a single conditional expectation, and we demonstrate how our results may be used for valid inference in this case. Such estimands naturally arise as identification formulae for causal estimands. The RDD setting is natural for our methodology since inference is important and linear smoothers are the natural second step regression method. En passant, we provide an extension of the RDD method to nonparametric methods for censored data as discussed in Remark~\ref{remark:RDD}.

\subsection{Background} \label{subsec:ApplicBackground}

The EMA program was established in England to encourage YP to continue their education after Year 11. It was piloted in September 1999, rolled out nationally in 2004, and abolished in September 2010. YP could apply for EMA during Year 11 and if EMA was awarded, YP would receive a weekly cash transfer during Year 12 and Year 13 provided they stayed in further education. EMA was awarded based on a household's annual income for the previous year submitted to the EMA administration via a bank statement. YP in households with annual incomes below £20,817 received £30, those between £20,818 and £25,521 received £20, and those between £25,522 and £30,810 received £10. No EMA was given for incomes over £30,810. The presence of these thresholds suggests that the causal effect of EMA can be estimated using an RDD. An RDD estimates a local causal effect by comparing groups just above and below a treatment threshold mimicking a (local) randomized controlled trial, see for example~\citet{Hahn.etal:2001},~\citet{Imbens:Lemieux:2008}, and~\citet{Cattaneo:Titiunik:2022}. An RDD is therefore able to infer causal effects under relatively weak assumptions, avoiding no unmeasured confounding and similar graphical causal model based criteria, confer with~\citet{Pearl:2009} and~\citet{Hernan:Robins:2020}. Previous analyses of the EMA data have relied on the assumption of no unmeasured confounding or have had methodological weaknesses, such as discarding censored observations and using polynomial regression rather than local linear regression. A detailed review of this prior literature is provided in Section D of the supplementary material.

We restrict our attention to measuring the effect of receiving high EMA since its effect is expected to be the highest and since 80\% of YP receiving EMA were paid the highest rate of £30, see~\citet{Bolton:2011}. Note that those not receiving high EMA could still be receiving moderate or low rates, and the causal effect estimated here is therefore only valid in environments where these rates are also present. Under assumptions about how CATE changes as a function of salary, e.g., linear dependence, one could exploit the multiple thresholds to infer the causal effect of high EMA versus no EMA but this is not pursued here. For simplicity, we similarly restrict attention to whether high EMA is received in Wave 4 making treatment binary. It would be of interest to extend this approach to dynamical treatments, using the treatment status from both Wave 4 and Wave 5.

\subsection{Model and results} \label{subsec:ApplicModelRes}

The present RDD is fuzzy since not all eligible YP apply for EMA and since the income information in LSYPE could deviate from the one submitted to the EMA administration. Additionally, the exact income is only available in Wave 1 and Wave 2 and in banded form in Wave 3 which was the year where EMA application were submitted. The income in Wave 3 is thus estimated by taking the income from Wave 2 if this is within the band and otherwise simulate uniformly over the band. Fortunately, the bands align well with the EMA thresholds, so the risk of moving an observation across a threshold is very low. A handful of seeds were tested for the simulation and they all gave quantitatively similar results in terms of the final estimate. 

Let $t=0$ be Wave 3, and the outcome $Y$ be the amount of years spent in full-time education during Wave 4 and Wave 5. Censoring $C$ takes the value $1$ if YP becomes censored in Wave 4, $2$ if YP becomes censored in Wave 5, and $3$ if not censored in Wave 4 and 5.  Let $X$ be baseline covariates from Waves 1-3 as well as a time-dependent coordinate which at the end of the year increases by 1 if YP was in full-time education during that school year so that $X^C$ is observable from the data and $Y=Y(X)$. Similarly, let the treatment outcome be denoted $A$ and $Z=(Z(t))_{t \geq 0}$ be as $X$ but where the time-dependent coordinate is 1 if YP receives high EMA at the end of the year and 0 otherwise such that $A=A(Z)$. Assume $(C,X^C)$ is a CAR of $X$ and $(C,Z^C)$ is a CAR of $Z$ and that positivity holds. Let $W$ be income in Wave 3 and $w_0=\textnormal{£20,817}$. Let $Y^{(a)}$ be the potential outcome corresponding to treatment $a \in \{0,1\}$ and specify the causal estimand of interest as the CATE
$$\tau = \mathbb{E}[Y^{(1)}-Y^{(0)} \mid W=w_0].$$ 
For identification, assumptions analogous to those in Theorem 2 of~\citet{Hahn.etal:2001} are imposed.

\begin{assumption} (RDD identification.) \label{assumption:RDD}
    \begin{enumerate}[label=(\roman*)]
    \item $a^+ = \lim_{w \downarrow w_0} \mathbb{E}[ A \mid W = w]$ and $a^-=\lim_{w \uparrow w_0} \mathbb{P}(A=1 \mid W = w)$ exist and $a^+ \neq a^-$.
    \item $\mathbb{E}[Y^{(1)} \mid W=w]$ and $\mathbb{E}[Y^{(0)} \mid W=w]$ are continuous in $w$ at $w_0$.
    \item $A \indep (Y^{(1)}-Y^{(0)}) \mid W=w$ in the limit for $w \rightarrow w_0$. \demoas
    \end{enumerate}
\end{assumption}
\noindent Theorem 2 in~\citet{Hahn.etal:2001} then implies 
$$\tau = \frac{y^+-y^-}{a^+-a^-}$$
where $y^+ = \lim_{w \downarrow w_0} \mathbb{E}[Y \mid W=w]$ and  $y^- = \lim_{w \uparrow w_0} \mathbb{E}[Y \mid W=w]$. Estimation of the four components $(y^+, y^-, a^+, a^-)$ proceeds by component-wise applying Algorithm~\ref{alg:DR} with linear smoothers. Assume oracle efficiency is obtained for each component, which holds under conditions given in Proposition~\ref{prop:asymp} and~\ref{prop:crossFit}. Then by Remark~\ref{rmk:VectorValued}, oracle efficiency is obtained for $\tau$ since this is determined by the asymptotic distribution of $(\hat{y}^+,\hat{y}^-,\hat{a}^+,\hat{a}^-)$, confer with~\citet{Hahn.etal:1999}. Standard implementations for inference based on asymptotic approximations may thus be used, treating the pseudo-outcomes as ordinary outcomes. 

Computation of the pseudo-outcomes can be formulated as sequential classification problems. Estimation is performed using the \texttt{R}-package \texttt{xgboost} where five-fold cross-validation with negative log-likelihood and AUC loss functions were used to determine suitable hyperparameters. Since \texttt{xgboost} is tree-based, it should be able to capture discontinuities caused by the EMA thresholds well. For predicting censoring, even with a moderate amount of hyperparameter optimization, it is hard to improve the performance of the model using only the covariates identified as predictors of non-response in Section 4.4 of~\citet{Collingwood.etal:2010} compared to using all available covariates. A high-dimensional $X$ may sometimes be desirable to make CAR more plausible, but since this does not seem to be needed here, we proceed with the lower dimensional model containing 14 covariates, even though the performance of the higher dimensional model is substantially better for predicting education outcomes. The analysis was also performed for the high-dimensional choice of $X$, but is not reported as it led to highly similar results, likely bechause of double robustness. The second-step estimator is a local-linear-regression-based RDD implemented using the \texttt{R}-package \texttt{rdrobust} with standard parameters except for the bandwidth, which is set to $h=3\,500$. This bandwidth was chosen to be slightly smaller than the distance to the next EMA threshold, with the aim of balancing the sample size and the homogeneity of the subjects. This leads to around $860$ and $720$ observations to the left and right of the threshold, respectively. Other bandwidth choices are discussed later.  It is an attractive feature of the approach that $X$ can be made high-dimensional to make CAR more plausible while keeping the final estimation as a low-dimensional local linear regression, which has desirable properties for inference. 

\begin{figure}[t]
    \centering
    \includegraphics[width=1\linewidth]{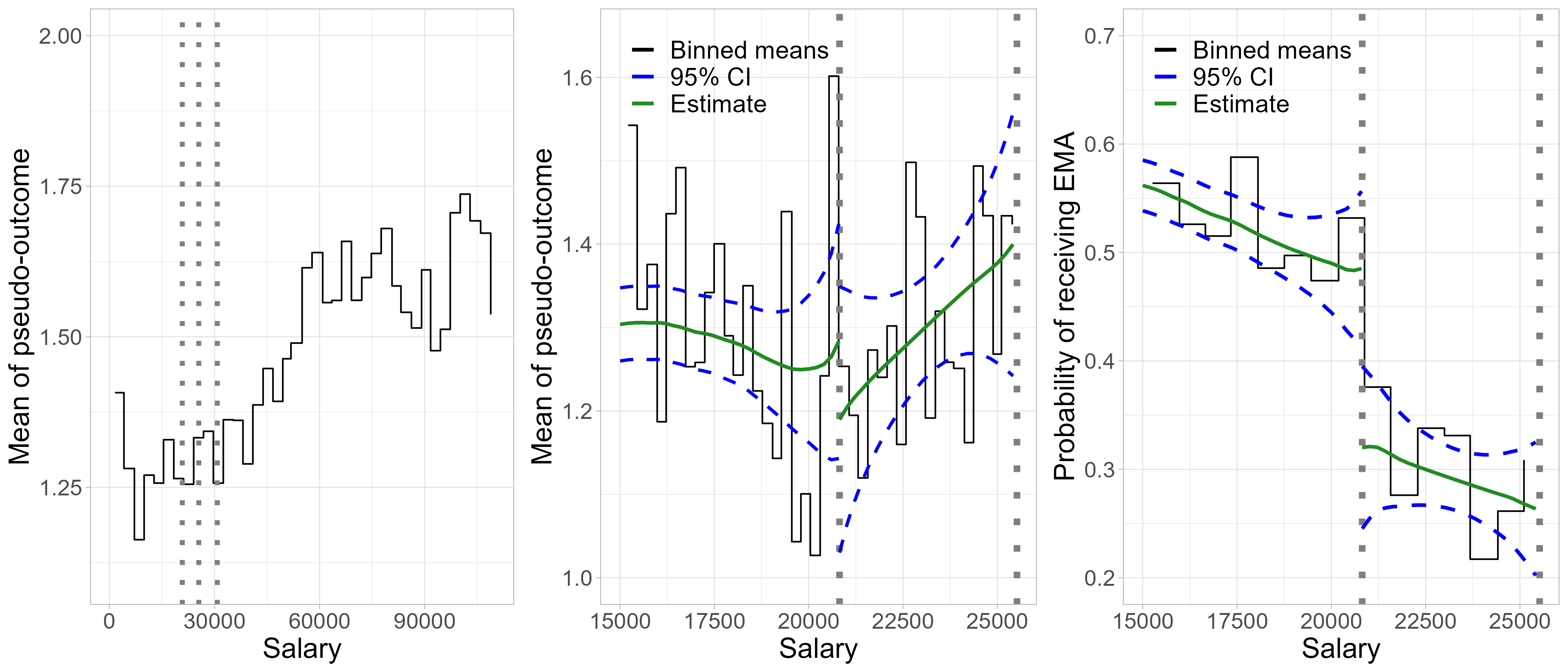}
    \caption{\textbf{Left Panel:} Binned-means over the pseudo-outcomes for $Y$ (black solid line) and EMA thresholds (gray dotted line) for income in Wave 3. \textbf{Middle Panel:} Binned-means of the pseudo-outcomes for $Y$ (black solid line), estimate of the conditional mean (green solid line), 95\% confidence intervals (blue dashed line), and EMA thresholds (gray dotted line) for Wave 3 incomes around the threshold for receiving high EMA. \textbf{Right Panel:} The same as the middle panel with pseudo-outcomes for $A$.}
    \label{fig:RDDplot}
\end{figure}

The relationship between the estimated pseudo-outcomes and the income in Wave 3 is depicted in the right panel of Figure~\ref{fig:RDDplot}. This shows a clear discontinuity in treatment probability at $w_0$ indicating that an RDD is indeed applicable. In the left-panel, one sees that the expected outcome seems to increase with salary until around \textnormal{£60,000} after which it appears constant. The level also appears constant until around \textnormal{£30,000} which could be an indication that the level below \textnormal{£20,000} is artificially high due to EMA. The middle panel focuses on a neighborhood of $w_0$, and also seems to indicate a discontinuity for the education outcomes although its statistical significance is less clear. A desirable feature of using pseudo-outcomes for RDD is that the regression discontinuity can still be plotted when data is censored. Such graphical tools are important for RDD analyses, see~\citet{Imbens:Lemieux:2008}.

\noindent Algorithm~\ref{alg:DR} leads to the estimated value and standard error
\begin{align*}
    \hat{\tau} = 0.703, \hspace{0.5cm} \textnormal{SE}(\hat{\tau})=0.614, 
\end{align*}
resulting in a $p$-value of around 0.25 and thus not reaching statistical significance at conventional significance levels. 

To get a feeling for the sensitivity of the result with respect to the bandwidth, the estimation was repeated with $h=4\,704$ and $h=2\,352$ which is the distance to the next EMA threshold and half that distance, respectively. The estimated values are in this case $\{\hat{\tau},\textnormal{SE}(\hat{\tau})\} = (0.391,0.588)$ and $\{\hat{\tau},\textnormal{SE}(\hat{\tau})\} = (1.197,0.697)$, respectively. Thus, the absolute size of the estimate changes considerably, but the effect remains large and positive. Note that the $90\%$ significance level is reached for the smaller bandwidth. Making Figure~\ref{fig:RDDplot} with these alternative bandwidths (not shown) indicates over- and undersmooth, respectively, and the original estimate hence seems to be the most reliable.

The proposed methods yield wider confidence intervals than those in~\citet{Rahman:2014}, leading to statistically insignificant results. This likely better reflects the uncertainty in the estimate since the assumptions imposed here are substantially weaker. If the effect is genuine, an additional $0.7$ years of education would be a large effect and would be another piece of evidence that EMA was successful in its initial aim of keeping YP in education. Transparency regarding uncertainty is important for policymakers when assessing findings and deciding if more data is needed before making decisions. The $p$-value suggests a larger sample could be beneficial in clarifying the effect, or statistical power could be increased by using data from the multiple EMA thresholds, as discussed, though at the cost of some internal validity. This is left for future work. Some possible extensions of the present analysis are discussed in Section D of the supplementary material.

\begin{remark} (RDD with survival data.) \label{remark:RDD} \\
The use of RDD for survival data has been studied in~\citet{Adeleke.etal:2022} for an accelerated failure time model. The methods proposed in this paper seem to be the first that allow for nonparametric inference for an RDD when data is censored even for the survival setting. Note that the outcome $Y$ specified above cannot be represented as survival data since some leave school in Wave 4 but return in Wave 5. 
    \demormk
    
\end{remark}

\section{Discussion}

This paper has introduced a method for doubly robust learning of regression functions with right-censored data using pseudo-outcomes. We have demonstrated that when cross-fitting is employed and the second-step regression method is a linear smoother, the approach inherits the asymptotic distribution of the linear smoother, provided the nuisance estimators converge at sufficient rates. This offers a robust and flexible pathway for making pointwise inference on conditional expectations in the presence of right-censoring. While we conjecture that the oracle efficiency results hold for a broader class of learners beyond linear smoothers, this remains an open question. It is anyhow generally difficult to establish the asymptotic distribution of nonparametric regression methods outside of the class of linear smoothers even in the uncensored setting. The proposed approach thus inherits these well-known challenges of nonparametric inference. Future work could also address the optimality of the estimator's minimax risk constant under cross-fitting and local linear smoothing, as well as investigate sufficient conditions for assumptions (ii), (iv), and (v) listed in Proposition 1.

The data application demonstrated the utility of the framework by constructing a novel nonparametric RDD estimator for right-censored data. This could be a valuable tool in exploring long-ranging consequences of policies in cases where a longitudinal no unmeasured confounding assumption might be unsuitable, but CAR for the censoring mechanism is believable. A limitation of this methodology, as encountered in our empirical analysis, is that the combination of data censoring and the local nature of RDD estimation necessitates large sample sizes to achieve adequate statistical power.

\section*{Acknowledgments}
Significant parts of the research were conducted during a visit to the Department of Mathematical Statistics at Stockholm University. The author gratefully acknowledges the hospitality of Filip Lindskog and Mathias Lindholm and thanks them for many fruitful discussions. The author also thanks their PhD supervisor Christian Furrer for general feedback and Niels Richard Hansen for helpful discussions. Finally, the author thanks the two anonymous reviewers for their careful reading and constructive comments, which significantly improved the quality of the paper.

\section*{Funding}
Oliver Lunding Sandqvist's research has partly been funded by the Innovation Fund Denmark (IFD) under File No.\ 1044-00144B.

\section*{Disclosure Statement}
The authors declare no conflicts of interest.

\section*{Supplementary Material}
\label{SM}
The supplementary material contains further discussions and the proofs.

\section*{Data Availability Statement}
The data that support the findings of this study are available at \url{ https://doi.org/10.5255/UKDA-SN-5545-7}. Restrictions apply to the availability of these data, which were used under license for this study.

\end{document}


\def\spacingset#1{\renewcommand{\baselinestretch}%
{#1}\small\normalsize} \spacingset{1}

\markboth{O.L.~Sandqvist}{A doubly robust learner for regression and inference with right-censored outcomes}

\title{Supplementary material for “A doubly robust learner for regression and inference with right-censored outcomes”}

  \author{O.L.~Sandqvist$^{\textnormal{a,b}}$ \vspace{0.5cm}\\
    $^\textnormal{a}$PFA Pension, Sundkrogsgade 4, DK-2100 Copenhagen \O, Denmark \\
    $^\textnormal{b}$Department of Mathematical Sciences, University of Copenhagen, \\ Universitetsparken 5, DK-2100 Copenhagen \O, Denmark} 

\maketitle

The supplementary material is organized as follows. Section~\ref{sec:additionalDiscussions} provides additional discussions and model extensions, Section~\ref{sec:AppendixEIF} derives the efficient influence function, Section~\ref{sec:AppendixSimulation} provides a detailed description of the data generating mechanism of the simulation study and the computation of the DRCUT, Section~\ref{sec:AppendixApplication} contains more background and possible extensions of the data application. Sections~\ref{sec:AppendixProof1},~\ref{sec:AppendixProof2}, and~\ref{sec:AppendixProof3} contains the proofs.
\vfill

\newpage
\spacingset{1.2}

\section{Additional discussions} \label{sec:additionalDiscussions}

\subsection{Jackknife pseudo-outcomes}
Pseudo-outcomes which are based on the jackknife rather than the efficient influence function have also been explored for survival and competing risks data. The jackknife pseudo-outcomes were introduced in~\citet{Andersen.etal:2003} and is an area of continued study, see for example~\citet{Jacobsen:Martinussen:2016},~\citet{Andersen.etal:2017},~\citet{Overgaard.etal:2017}, and~\citet{Parner.etal:2023}. The latter paper proposes to use so-called infinitesimal jackknife pseudo-outcomes, which turn out to be exactly the uncentered efficient influence function for $\mathbb{E}[Y]$. The motivation in that paper is that the jack-knife pseudo-outcomes were observed to be asymptotically equivalent to the infinitesimal jackknife pseudo-outcomes under suitable regularity conditions, and the latter were found to be much faster to compute. Infinitesimal jackknife pseudo-outcomes and DRCUT pseudo-outcomes are thus identical, which seems to have been overlooked by~\citet{Parner.etal:2023}. It is however fruitful to make this connection; an assumption that has been persistent in all literature on jackknife pseudo-outcomes is that the censoring is completely random $C \indep (T,W)$. This has been highlighted as a key assumption, but also as perhaps the most restrictive one, see e.g.\ Section 4 in~\citet{Overgaard.etal:2017}. Instead of taking the jackknife pseudo-outcomes as a starting point, the DRCUT, or equivalently the infinitesimal jackknife pseudo-outcomes, is here taken as the starting point. With this point of view, it is not difficult to allow the censoring distribution to depend on covariates and the past trajectory of $X$ since the efficient influence function is still well-known in this situation. Thus, this key independence assumption is relaxed substantially. 

\subsection{Estimands with dynamical conditioning information}
 It is possible to use the proof of Theorem~\ref{thm:DRCUT} to show that the transformation
    \begin{align*}
        Y^\ast_{\mathbb{P}_1,\mathbb{P}_2}(t,C,X^C) &= \frac{Y(X) 1_{(C \geq \eta)}}{\mathbb{P}_1(C \geq \eta \mid X)} \\
        &+ \int_{[0,\eta)} \frac{\mathbb{E}_2[Y(X) \mid X^{u \vee t}]}{\mathbb{P}_1(C > u \mid X)} \left(\diff 1_{(C \leq u)} - 1_{(C \geq u)} \frac{\mathbb{P}_1(C \in \diff u \mid X)}{\mathbb{P}_1(C \geq u \mid X)} \right)
    \end{align*}
    for any $t \geq 0$ satisfies
    \begin{align*}
        &\mathbb{E}[Y^\ast_{\mathbb{P}_1,\mathbb{P}_2}(t,C,X^C) - Y(X) \mid X^t] \\
        &= \mathbb{E}\left[ \int_{[0,\eta)} \mathbb{E}[Y(X) \mid X^{u \vee t}] - \mathbb{E}_2[Y(X) \mid X^{u \vee t}] \; \diff \left( \frac{\mathbb{P}(C > u \mid X)}{\mathbb{P}_1(C > u \mid X)} \right) \mid X^t \right].
    \end{align*}
    This extends the use of doubly robust transformations to parameters on the form $\mathbb{E}[Y(X) \hspace{-0.1cm} \mid X^t]$ which are often of interest. An example could be the total duration spent as disabled in an illness-death model given the state and duration at time $t$. The methods and results of this paper are straightforward to generalize to such estimands.   

    Allowing for dependence on the past trajectory of $X$ has been trivial in the cases considered in the literature so far since the censored data was always competing risks data; the paper~\citet{Sabathe.etal:2020} would appear to be an exception since they work in an illness-death model, but due to the estimands they consider, the setup could equivalently have been formulated as a competing risks model. A consequence of allowing for the dependence on covariates and the past trajectory for $X$ is that the nuisance parameters become more complicated. When nuisance estimators are simple, such that they do not overfit the data (e.g., if they belong to a Donsker class), they may usually be estimated in-sample without affecting the asymptotic distribution of a second-step estimator. This, for example, holds when the estimators are sufficiently smooth functions of sample averages, which is the case that has been explored in the jackknife pseudo-outcome literature so far, see for example~\citet{Overgaard.etal:2017} and~\citet{Parner.etal:2023}. The $K$-fold cross-fitting approach proposed in this paper allows for valid inference in the presence of flexible nuisance estimators, e.g., depending on hyperparameters selected in data-adaptive ways.

\subsection{Estimands with counterfactual outcomes} \label{subsec:causalEstimand}

Let $A$ be a coordinate of $X$ denoting the observed treatment which is assumed to be binary. Let $X^{(a)}$ be the potential outcome corresponding to what would have happened if $A$ had been $a \in \{0,1\}$ and set $X=X^{(A)}$. Assume no unmeasured confounding $A \indep X^{(a)} \mid W$ which leads to the following identification formula for the treatment-specific conditional mean
    \begin{align*}
        \mathbb{E}[Y(X^{(a)}) \mid W] = \mathbb{E}[Y(X^{(a)}) \mid W, A=a] = \mathbb{E}[Y(X) \mid W, A=a].
    \end{align*}
    We also impose treatment positivity $\mathbb{P}(A=a \mid W) \geq \epsilon > 0$. 

The efficient influence function can now be derived using similar arguments to those in Section~\ref{sec:AppendixEIF}. Define the inverse probability weighted pseudo-outcomes for treatment $a$ as
    $$Y^\circ(a,C,X^C) = \frac{Y(X)1_{(C \geq \eta)}1_{(A=a)}}{\mathbb{P}(C \geq \eta \mid X) \mathbb{P}(A=a \mid W)}.$$
    Sample means of these pseudo-outcomes provide an estimator of $\mathbb{E}[ \mathbb{E}[Y(X) \mid W, A=a] ]$, which is the population version of the estimand of interest $\mathbb{E}[Y(X) \mid W, A=a]$. For these estimands, $(C,X^{(a)},A)$ is the complete data and $(C,X^C)$ is the observed data. Following the arguments in~\citet{Rytgaard.etal:2022}, the projection onto the relevant tangent space is given by 
    \begin{align*}
        &\int_{[0,\eta)} \mathbb{E}[Y^\circ(a,u,X^C) \mid C > u, X^u] - \mathbb{E}[Y^\circ(a,C,X^C) \mid C > u, X^u] \; M(\diff u) \\
        &+\sum_{k \in \{0,1\}} \left( \mathbb{E}[Y^\circ(a,C,X^C) \mid W, A=k] - \mathbb{E}[Y^\circ(a,C,X^C) \mid W]  \right) 1_{(A=k)} \\
        &= -\frac{1_{(A=a)}}{\mathbb{P}(A=a \mid W)} \int_{[0,\eta)} \hspace{-0.1cm}\frac{\mathbb{E}[Y(X) \mid X^u]}{\mathbb{P}(C > u \mid X)} M(\diff u) +\frac{1_{(A=a)}- \mathbb{P}(A=a \mid W)}{\mathbb{P}(A=a \mid W)} \mathbb{E}[Y(X) \mid W, A=a]
    \end{align*}
    with $M$ defined as in Appendix~\ref{sec:AppendixEIF}. This result also appears in Section 6.4.3 of~\citet{Van:Robins:2003} when $Y(X)=1_{(T \leq t)}$ for a survival time $T$.

Hence, the uncentered efficient influence function motivates the transformation
    \begin{align*}
    Y^\ast_{\mathbb{P}_1,\mathbb{P}_2,\mathbb{P}_3}(a,C,X^C) &= \frac{Y(X)1_{(C \geq \eta)}1_{(A=a)}}{\mathbb{P}_1(C \geq \eta \mid X) \mathbb{P}_3(A=a \mid W)} \\
        &+\frac{1_{(A=a)}}{\mathbb{P}_3(A=a \mid W)} \hspace{-0.1cm} \int_{[0,\eta)} \hspace{-0.125cm}\frac{\mathbb{E}_2[Y(X) \mid X^u]}{\mathbb{P}_1(C > u \mid X)} \left(\diff 1_{(C \leq u)} \hspace{-0.1cm} - \hspace{-0.05cm} 1_{(C \geq u)} \frac{\mathbb{P}_1(C \in \diff u \mid \hspace{-0.1cm} X)}{\mathbb{P}_1(C \geq u \mid \hspace{-0.1cm} X)} \right) \\
        &-\frac{1_{(A=a)}- \mathbb{P}_3(A=a \mid W)}{\mathbb{P}_3(A=a \mid W)} \mathbb{E}_2[Y(X) \mid W, A=a] \\
        &= \frac{1_{(A=a)}}{\mathbb{P}_3(A=a \mid W)} \left(  Y_{\mathbb{P}_1,\mathbb{P}_2}^\ast(C,X^C) - \mathbb{E}_2[Y(X)\mid W,A=a] \right) \\
        &+ \mathbb{E}_2[Y(X) \mid W,A=a].
    \end{align*}
    Related results are Theorem 6.1 in~\citet{Van:Robins:2003} for the discrete-time case and Theorem 1 in~\citet{Rytgaard.etal:2022} for the continuous-time case where treatment starts strictly after time $0$. It is worth noting that the final expression is the same as the usual efficient influence function in the absence of censoring, except that the outcomes $Y$ have been replaced by the pseudo-outcomes $Y_{\mathbb{P}_1,\mathbb{P}_2}^\ast(C,X^C)$. By calculations similar to those in the proof of Theorem~\ref{thm:DRCUT}, one can show $\mathbb{E}[ Y^\ast_{\mathbb{P},\mathbb{P},\mathbb{P}}(a,C,X^C) \mid W] = \mathbb{E}[Y(X) \mid W, A=a]$. This implies
    \begin{align*}
        &\mathbb{E}[Y^\ast_{\mathbb{P}_1,\mathbb{P}_2,\mathbb{P}_3}(a,C,X^C) - Y^\ast_{\mathbb{P},\mathbb{P},\mathbb{P}}(a,C,X^C) \mid W] \\
        &= \mathbb{E}\left[ Y^\ast_{\mathbb{P}_1,\mathbb{P}_2,\mathbb{P}_3}(a,C,X^C) - \frac{Y(X)1_{(A=a)}}{\mathbb{P}_3(A=a \mid W)}  - \left(\frac{1_{(A=a)}}{\mathbb{P}(A=a \mid W)} - \frac{1_{(A=a)}}{\mathbb{P}_3(A=a \mid W)} \right) Y(X)  \mid W \right] \\
        &= \mathbb{E}\bigg[ \frac{1_{(A=a)}}{\mathbb{P}_3(A=a \mid W)} \int_{[0,\eta)} \mathbb{E}[Y(X) \mid X^u] - \mathbb{E}_2[Y(X) \mid X^u] \; \diff \left( \frac{\mathbb{P}(C > u \mid X)}{\mathbb{P}_1(C > u \mid X)} \right) \mid W\bigg]  \\
        & \qquad +\frac{\mathbb{P}(A=a \mid W) - \mathbb{P}_3(A=a \mid W)}{\mathbb{P}_3(A=a \mid W)} \left( \mathbb{E}[Y(X) \mid W, A=a]-\mathbb{E}_2[Y(X) \mid W, A=a] \right) 
    \end{align*}
    where the last equality follows from factoring out $1_{(A=a)}/\mathbb{P}_3(A=a \mid W)$ in the first two terms of the transformation and then proceeding as in Theorem~\ref{thm:DRCUT}. This is analogous to Theorem~\ref{thm:DRCUT} and implies that the conditional mean is correct if either $\mathbb{P}_2=\mathbb{P}$ or $\mathbb{P}_1=\mathbb{P}_3=\mathbb{P}$. 
    The methods and results of this paper generalize straightforwardly to such estimands.

\subsection{Other extensions}

It may be possible to use alternative influence function pseudo-outcomes. As noted in~\citet{Young:Shah:2024}, the efficient influence function usually requires estimation of complex objects that can be ill-behaved in finite samples (e.g., inverse probability weights). For the partially linear model,~\citet{Young:Shah:2024} identifies a collection of influence functions, containing the efficient influence function, that retain desirable theoretic properties. They then select the one that minimizes the empirical sandwich variance. These ideas may be transferable to the censored data setting, and may provide alternative influence function pseudo-outcomes.

A related extension would be to use efficient influence function pseudo outcomes for model specifications that target covariates other than the censoring variable (as in the main text) and the treatment indicator (as in Section~\ref{subsec:causalEstimand}). One such example is the partially linear model where $X=(W,Z,\varepsilon)$ and
$$Y(X) = W^T \theta + f(Z)+\varepsilon$$
for the inferential target $\theta$ and errors satisfying $\mathbb{E}[\varepsilon \mid W,Z]=0$. The efficient influence function for $\theta$ targets the effect of $W$ on $Y$, and could likely be used as pseudo-outcomes in standard regression  to obtain doubly-robust and efficient estimation.

\subsection{Doubly robust random forests}

\citet{Steingrimsson.etal:2016} and~\citet{Steingrimsson.etal:2019} use a DRCUT for the composite outcome $Y(X)=L(T,W)$ where $L$ is a loss function. This is sufficient for their purposes since the second-step estimator $\hat{\mathbb{E}}_n$ is restricted to regression trees and random forests, see Algorithm 1 and 2 of~\citet{Steingrimsson.etal:2019}, which only depend on data through the loss of the individual observations. Differently from Algorithm~\ref{alg:DR}, it is proposed to estimate the nuisance parameters using all the data, and then to fit the regression model on the full data set of pseudo-outcomes. 
    
In Section 2.4 of~\citet{Steingrimsson.etal:2016}, the possibility of estimating nuisance parameters in parallel with fitting the regression trees is discussed and it is stated to impair performance. The possibility of using sample splitting is not discussed. On the contrary,~\citet{Steingrimsson.etal:2016} states that "...it is not obvious why using pre-computed estimators of these functions derived from the entire dataset should lead to overly optimistic risk estimators". Even if the approach does not lead to overfitting when the goal is prediction, a nuisance estimation performed in-sample can affect inference due to the added variability induced by estimating the nuisance parameters, see for example~\citet{Chen.etal:2003}. See however the discussion in Section~\ref{subsec:noSplitting}. For valid inference, one would need to quantify this added variability and adjust the standard errors coming from the second-step regression accordingly. When nuisance estimation is performed using sample-splitting, Proposition~\ref{prop:asymp} implies that the effect on inference is simple; the estimator behaves as if one had access to the oracle pseudo-outcomes but only had $n$ observations instead of $2n$. Cross-fitting may be used to regain full-sample efficiency.  

\subsection{Mean squared error convergence rates}

 In some situations, it might be more natural to take the convergence rate of the oracle mean squared error $R^\ast_n(w) = \mathbb{E}[\{\tilde{m}(w)-m(w) \}^2]^{1/2}$ as a starting point rather than that of $\tilde{m}(w)-m(w)$. Straightforward calculations show that if $n^\alpha \{\Tilde{m}(w)-m(w)\}$ converges in distribution to some distribution with mean $\mu$ and variance $\sigma^2$ and the first and second moments also converge then $n^\alpha R^\ast_n(w) \rightarrow (\sigma^2+\mu^2)^{1/2}$ for $n \rightarrow \infty$ implying $R^\ast_n(w) = O_{\mathbb{P}}(n^{-\alpha})$. In other words, studying the pointwise convergence rates is equivalent to studying the convergence rate of the mean squared error. Hence, if the integrated MSE rather than the pointwise MSE is the relevant performance metric, the approach in~\citet{Rambachan.etal:2022} might be more suitable.

\subsection{Oracle efficiency without sample splitting} \label{subsec:noSplitting}

If nuisance estimators converge sufficiently fast and uniformly, the added variance from estimating nuisance parameters in-sample may become asymptotically negligible, see for example Lemma 19.24 in~\citet{Van:1998} and Lemma 2 in~\citet{Cui.etal:2023}. In this case, sample-splitting is not necessary and the conditional bias is less relevant. Furthermore, that approach might generalize more easily to estimators that are not linear smoothers. The present approach is chosen to allow for weaker conditions on the convergence rates which exploit the product structure of the conditional bias. 

\subsection{Bias-variance tradeoff with cross-fitting}

It can be seen from the proof of Proposition~\ref{prop:crossFit} that the asymptotic distribution comes from averaging $\tilde{m}_k(w)$ ($k=1,\dots,K$). We therefore conjecture that a DML2-variant would, under similar regularity assumptions, have the same asymptotic distribution as $\tilde{m}(w)$. Thus, when $\alpha < 1/2$, the estimator $\hat{m}^{\textnormal{CF}}(w)$ trades a decrease in variance for an increase in bias compared to a DML2-variant. This phenomenon seems to imply that an estimator converging at an optimal rate (in the sense of~\citet{Stone:1980,Stone:1982}) which is slower than $\sqrt{n}$ cannot have a non-zero asymptotic bias since splitting and averaging would then decrease the asymptotic variance without a corresponding increase in bias. For an asymptotically unbiased estimator converging at a sub-optimal rate, e.g., univariate local linear regression with undersmoothing, the variance reduction due to averaging is analogous to increasing the proportionality constant in the bandwidth and can hence be thought of as smoothing.

\section{Derivation of the efficient influence function} \label{sec:AppendixEIF}

    Sample means of the IPCW pseudo-outcomes 
     $$Y^\circ(C,X^C) = \frac{Y(X)1_{(C \geq \eta)}}{\mathbb{P}(C \geq \eta \mid X)}$$ 
     provide an estimator of $\mathbb{E}[Y(X)]$. Furthermore, this estimator is a regular and linear (hence also asymptotically linear) estimator of $\mathbb{E}[Y(X)]$ with influence function $Y^\circ(C,X^C)-\mathbb{E}[Y(X)]$, see e.g., Section 3 in~\citet{Tsiatis:2006}. The efficient influence function is any influence function subtracted its projection in $L^2(\Omega,\mathcal{F},\mathbb{P})$ onto the CAR tangent space which may be found using~\citet{Van:2004} or Section 3.4 of~\citet{Van:Robins:2003}. Let $M(\diff u) = \diff 1_{(C \leq u)} - 1_{(C \geq u)} \mathbb{P}_1(C \in \diff u \mid X)/\mathbb{P}_1(C \geq u \mid X)$. The aforementioned projection is then 
     $$\int \mathbb{E}[Y^\circ(u,X^u)-\mathbb{E}[Y(X)] \mid C>u, X^u ]-\mathbb{E}[Y^\circ(C,X^C)-\mathbb{E}[Y(X)] \mid C>u, X^u] \; M(\diff u)$$
    when the conditional expectations are taken to be $0$ if $\mathbb{P}(C > u \mid X^u)=0$. Thus, there are only contributions on $[0,\eta)$. The marginals expectations cancel and since $1_{(u \geq \eta)}=0$ for $u \in [0,\eta)$ the only remaining term is $-\int_{[0,\eta)} \mathbb{E}[ Y^\circ(C,X^C) \mid C>u, X^u] \; M(\diff u)$.
    Note
    \begin{align*}
        \mathbb{E}[Y^\circ(C,X^C) \mid C>u, X^u]  &= \frac{\mathbb{E}[Y(X)1_{(C \geq \eta)}/\mathbb{P}(C \geq \eta \mid X) \mid X^u] }{\mathbb{P}(C > u \mid X^u)}
    \end{align*}
    since $1_{(C \geq \eta)} 1_{(C > u)} = 1_{(C \geq \eta)}$ for $u \in [0,\eta)$. Using the tower-property when conditioning on $X$ in the numerator gives $\mathbb{E}[Y(X) \mid X^u]/\mathbb{P}(C > u \mid X^u)$.

\section{Additional details for simulation study} \label{sec:AppendixSimulation}

For a given subject, the data is generated as follows: The hazard $\gamma$ of $C \mid X$ is set to $\gamma(t,W)=1\{Z(t) = 1\}\exp(\log(0.2)+0.6 \times  1\{-2 \leq W < 2\})$, which results in a substantial amount of right-censoring as well as highly state-dependent censoring. Subjects not censored before time $\eta$ are administratively censored.  Events are simulated according to the transition hazards
\begin{align*}
    \mu_{12}(t, W) &= \exp\big(\log(0.3)+0.15 \times \cos(\pi W /2)+0.15 \times 1\{ t > 2.5\}-0.05 \times W\big), \\
    \mu_{13}(t, W) &= \exp\big(\log(0.1)+0.3 \times \sin(\pi W/2)+0.05 \times t \big), \\
    \mu_{23}(t,S(t), W) &= \exp\big(-0.75 \times \min \{t-S(t),3\} \times (1.07+0.09 \times \bar{W}-0.024 \times \bar{W}^2 \\
    & \qquad \quad -0.014 \times \bar{W}^3+0.001 \times \bar{W}^4+0.00065 \times \bar{W}^5 )  \big),
\end{align*}
where $S(t) = \sup \{ s \leq t : Z(s) \neq Z(t) \}$ is the latest jump time and $\bar{W} = \min\{W,3\}$. With these specifications, Assumption~\ref{assumption:CAR} holds because the censoring intensity is adapted to the filtration generated by $X$ and Assumption~\ref{assumption:positivity} holds because the largest probability of becoming censored before time $\eta$ is obtained by remaining in the Healthy state, and this leads to a censoring probability that is strictly less than one. In addition, $Y$ clearly has finite expectation. The required assumptions for use of IPCW and doubly robust pseudo-outcomes are hence satisfied.

For computation of $Y^\ast$ it is convenient to introduce the prospective illness duration $Y(X,t)=\int_{(t,\eta)} 1\{Z(s)=2\} \diff s$ and its conditional expectation $V_{Z(t)}(t,S(t),W) = \mathbb{E}[Y(X,t) \mid X^t]$. Then 
$\mathbb{E}[Y(X) \mid X^t] = \int_{(0,t]} 1 \{ Z(s)=2\} \diff s + V_{Z(t)}(t,S(t),W)$
and $V$ may be calculated using the differential equation from Corollary 7.2 in~\citet{Adekambi:Christiansen:2017} whenever transition hazards exist, giving
\begin{align*}
    \frac{\diff}{\diff t} V_1(t,s,w) &= (\mu_{12}(t,w)+\mu_{13}(t,w)) \times V_1(t,s,w) - \mu_{12}(t,s,w) \times V_{2}(t,t,w), \\
    \frac{\diff}{\diff t} V_2(t,s,w) &= -1+\mu_{23}(t,s,w) \times V_2(t,s,w),
\end{align*}
with boundary conditions $V_1(\eta,s,w)=V_2(\eta,s,w)=0$. The fourth-order Runge-Kutta method is used to solve these differential equations. Note that computing $V_j(t,s,w)$ via this approach also yields $V_j(u,s,w)$ for all $u \geq t$.

\section{Additional details for data application} \label{sec:AppendixApplication}

\subsection{Background}

Studies based on self-reports indicated that "only" 12\% of recipients stayed in education because of EMA, which the government used as a key reason for abolishing EMA, see~\cite{Bolton:2011}. This highlights the importance of statistical analyses in evaluating the effectiveness of such programs to guide informed policymaking. These numbers were consistent with other studies that used matching between the pilot and control groups, see~\citet{Maguire.etal:2001},~\citet{Middleton.etal:2005}, and~\citet{Dearden.etal:2009}. An issue that was identified, but not controlled for, was that students staying in full-time education seemed more likely to remain in the survey, see e.g., Chapter 2.5.3 of~\citet{Middleton.etal:2005}. Additionally, the effect of EMA in the pilot might have been different than the national effect. 

Not many studies have explored the effect of EMA after it was rolled out nationally. The only studies identified on the topic were~\citet{Holford:2015}, the working paper~\citet{McKendrick:2022}, and the unpublished PhD~\citet{Rahman:2014} that employ panel regression, augmented inverse propensity weighted linear regression and Causal Forests, and an RDD, respectively. Except for the RDD, all previous studies hence rely on the assumption of no unmeasured confounding. The RDD in~\citet{Rahman:2014} had some methodological weaknesses which are improved upon in this analysis. Firstly, observations in Wave 4 and Wave 5 were pooled such that a YP interviewed in Wave 4 and Wave 5 would contribute with two observations. Censored observations were discarded. This can create confounding over time e.g., if YP that responded positively to EMA and stayed in education were more likely to respond to the survey as was found in~\citet{Middleton.etal:2005}. Secondly, polynomial regression was used to estimate the relevant conditional expectations and to perform inference. As noted in~\citet{Hahn.etal:1999}, this is fragile to misspecification, so local linear regression might be preferred since it is nonparametric and has good boundary properties.

Consequently, we find that an RDD based on observational data and utilizing the proposed methods can be a valuable complementary study for measuring the effect of EMA since it does not rely on no unmeasured confounding, allows the censoring distribution to depend on whether YP stays in education or not, uses the cohort is the one that emerged when EMA was well-established on a national level, and allows for the use of flexible nonparametric estimators for inference. This leads to both higher internal and external validity of the estimates.

\subsection{Extensions for LSYPE RDD}

A slightly more sophisticated model would have accommodated the fact that interviews took place over a few months rather than simultaneously, using that the interview month is available from the data to model $C$ on a monthly rather than yearly grid. The effect of this is, however, expected to be minor in the present study. Additionally, one could have weakened the assumption that $(C,X^C)$ is a CAR of $X$ by including more outcomes from Wave 4 in $X$, but nuisance estimators would then have to model the entire distribution of $X$ at Wave 4 given $X^0$ or use landmarking.

There are several long-term consequences of EMA that could be interesting to analyze using the LSYPE data. Waves 6-8 enable examination of university attendance and choice of subjects, and Wave 8 contains information on labour market outcomes. Other linked administrative data are also available, though under stricter access requirements. Similar datasets are however, available during COVID-19 years (2020-2021), allowing one to explore the long-term effect of EMA on self-reported health, amount of hours worked, trust in the government, etc. Here, it might be natural to let time $0$ be Wave 4, such that treatment is a baseline covariate and the approach from Section~\ref{subsec:causalEstimand} may be used. This is left to future work.

\section{Proof of Theorem~\ref{thm:DRCUT}} \label{sec:AppendixProof1}

The proof proceeds in two parts.

\subsection{Proof of conditional expectation result}

     \begin{proof}
 Write
\begin{align*}
    Y^\ast_{\mathbb{P}_1,\mathbb{P}_2}(C,X^C) &= \frac{Y(X) 1_{(C \geq \eta)}}{\mathbb{P}_1(C \geq \eta \mid X)} + \frac{\mathbb{E}_2[Y(X) \mid X^u]}{\mathbb{P}_1(C > u \mid X)} \Big\vert_{u=C} \times 1_{(C < \eta)} \\
    & \qquad - \int_{[0,\eta)} 1_{(C \geq u)} \frac{\mathbb{E}_2[Y(X) \mid X^u]}{\mathbb{P}_1(C > u \mid X)} \frac{\mathbb{P}_1(C \in \diff u \mid X)}{\mathbb{P}_1(C \geq u \mid X)}.
\end{align*}
The conditional expectation given $W$ of each term is treated separately. The strategy is to write the expectation in terms of $X \mid W$ and $C \mid X$. 
\begin{align*}
    \mathbb{E}\left[ \frac{Y(X) 1_{(C \geq \eta)}}{\mathbb{P}_1(C \geq \eta \mid X)} \mid W \right] &= \mathbb{E}\left[ \frac{Y(X)\mathbb{P}(C \geq \eta \mid X)}{\mathbb{P}_1(C \geq \eta \mid X)} \mid W \right]
\end{align*}
by the tower-property when conditioning on $X$. Similarly,
\begin{align*}
    \mathbb{E}\left[\frac{\mathbb{E}_2[Y(X) \mid X^u]}{\mathbb{P}_1(C > u \mid X)} \Big\vert_{u=C} \times 1_{(C < \eta)} \mid W\right] &= \mathbb{E}\left[ \int_{[0,\eta)} \frac{\mathbb{E}_2[Y(X) \mid X^u]}{\mathbb{P}_1(C > u \mid X)} \mathbb{P}(C \in \diff u \mid X) \mid W \right]
\end{align*}
and
\begin{align*}
    &\mathbb{E}\left[\int_{[0,\eta)} 1_{(C \geq u)} \frac{\mathbb{E}_2[Y(X) \mid X^u]}{\mathbb{P}_1(C > u \mid X)} \frac{\mathbb{P}_1(C \in \diff u \mid X)}{\mathbb{P}_1(C \geq u \mid X)} \mid W\right] \\
    &= \mathbb{E}\left[ \int_{[0,\eta)}  \frac{\mathbb{P}(C \geq u \mid X) \mathbb{E}_2[Y(X) \mid X^u]}{\mathbb{P}_1(C > u \mid X)} \frac{\mathbb{P}_1(C \in \diff u \mid X)}{\mathbb{P}_1(C \geq u \mid X)} \mid W \right].
\end{align*}
Note that
$$\frac{\mathbb{P}(C \in \diff u \mid X)}{\mathbb{P}_1(C > u \mid X)}=\gamma(u \mid X) \frac{\mathbb{P}(C \geq u \mid X)}{\mathbb{P}_1(C > u \mid X)} \diff \mu(u)$$
and
$$\frac{\mathbb{P}(C \geq u \mid X)}{\mathbb{P}_1(C > u \mid X)} \frac{\mathbb{P}_1(C \in \diff u \mid X)}{\mathbb{P}_1(C \geq u \mid X)} = \gamma_1(u \mid X) \frac{\mathbb{P}(C \geq u \mid X)}{\mathbb{P}_1(C > u \mid X)} \diff \mu(u).$$
Thus,
    \begin{align*}
        &\mathbb{E}[Y^\ast_{\mathbb{P}_1,\mathbb{P}_2}(C,X^C) - Y(X) \mid W] \\
        &= \mathbb{E}\bigg[ \frac{\mathbb{P}(C \geq \eta \mid X)}{\mathbb{P}_1(C \geq \eta \mid X)}Y(X) - Y(X) \\
        & \qquad + \int_{[0,\eta)} \mathbb{E}_2[Y(X) \mid X^u] (\gamma(u \mid X)-\gamma_1(u \mid X)) \frac{\mathbb{P}(C \geq u \mid X)}{\mathbb{P}_1(C > u \mid X)} \diff \mu(u)  \mid W\bigg].
    \end{align*}
Using p.\ 868 of~\citet{Shorack:Wellner:1986}, write
    \begin{align*}
        \frac{\mathbb{P}(C \geq \eta \mid X)}{\mathbb{P}_1(C \geq \eta \mid X)}Y(X) - Y(X) &= \int_{[0,\eta)} Y(X) \diff \left( \frac{\mathbb{P}(C > u \mid X)}{\mathbb{P}_1(C > u \mid X)} \right).
    \end{align*}
Integration by parts for finite variation functions, see p.\ 868 of~\citet{Shorack:Wellner:1986}, implies
    \begin{align*}
        \diff \left(\frac{\mathbb{P}(C > u \mid X)}{\mathbb{P}_1(C > u \mid X)} \right) = -\frac{\mathbb{P}(C \in \diff u \mid X)}{\mathbb{P}_1(C > u \mid X)} +  \frac{\mathbb{P}( C \geq u \mid X) }{\mathbb{P}_1(C \geq u \mid X) \mathbb{P}_1(C > u \mid X)} \mathbb{P}_1(C \in \diff u \mid X)
    \end{align*}
using Assumption~\ref{assumption:positivity} for $\mathbb{P}_1$. By the previous calculations, one therefore obtains
\begin{align*}
    \int_{[0,\eta)} Y(X) \diff \left( \frac{\mathbb{P}(C > u \mid X)}{\mathbb{P}_1(C > u \mid X)} \right) = -\int_{[0,\eta)} Y(X) (\gamma(u \mid X)-\gamma_1(u \mid X) ) \frac{\mathbb{P}(C \geq u \mid X)}{\mathbb{P}_1(C > u \mid X)} \diff \mu(u)
\end{align*}
\noindent Because of CAR, it holds that
\begin{align*}
    \mathbb{P}_1(C > u \mid X) = 1-\int_{[0,u]} r_1(s \mid X) \diff \mu(s)  = 1-\int_{[0,u]} \tilde{r}_1(s, X^s) \diff \mu(s)     
\end{align*}
so $\mathbb{P}_1(C > u \mid X) =  \mathbb{P}_1(C > u \mid X^u)$ by the tower property. Similar calculations hold for $\mathbb{P}_1(C \geq u \mid X)$ and $\mathbb{P}(C \geq u \mid X)$. Hence,
    \begin{align*}
        &\mathbb{E}\left[ \int_{[0,\eta)} Y(X) (\gamma(u \mid X)-\gamma_1(u \mid X) ) \frac{\mathbb{P}(C \geq u \mid X)}{\mathbb{P}_1(C > u \mid X)} \diff \mu(u) \mid W \right] \\
        &= \mathbb{E}\left[ \int_{[0,\eta)} \mathbb{E}[Y(X) \mid X^u] (\gamma(u \mid X)-\gamma_1(u \mid X) ) \frac{\mathbb{P}(C \geq u \mid X)}{\mathbb{P}_1(C > u \mid X)} \diff \mu(u) \mid W \right]
    \end{align*}
    using Fubini to take the expectation inside the integral, then tower with $X^u$ and use Fubini to take the expectation outside again. Collecting the results leads to the desired expression.
\end{proof}

\subsection{Proof of conditional variance result}

\begin{proof}
    For shorthand, write $Y^\ast = Y^\ast_{\mathbb{P},\mathbb{P}}(C,X^C)$ and $Y^\circ=Y^\circ(C,X^C)$ recalling the IPCW notation from Appendix~\ref{sec:AppendixEIF}. The first part of the proof generalizes the calculations from Proposition 5 of~\citet{Suzukawa:2004} and S.5.3 in the supplementary material of~\citet{Steingrimsson.etal:2019}. Note
    $$\Var[Y^\ast \mid W] = \mathbb{E}[(Y^\ast)^2 \mid W]-\mathbb{E}[Y^\ast \mid W]^2.$$ 
    By the first part of Theorem~\ref{thm:DRCUT}, it holds that $\mathbb{E}[Y^\ast \mid W]=\mathbb{E}[Y \mid W]$. For the other term, expanding the square gives $(Y^\ast)^2=R^{(1)}+R^{(2)}+R^{(3)}$ where
    \begin{align*}
        R^{(1)} &= (Y^\circ)^2, \\
        R^{(2)} &= \left(\int_{[0,\eta)} \frac{\mathbb{E}[Y \mid X^u]}{\mathbb{P}(C > u \mid X)} \left(\diff 1_{(C \leq u)} - 1_{(C \geq u)} \frac{\mathbb{P}(C \in \diff u \mid X)}{\mathbb{P}(C \geq u \mid X)} \right) \right)^2, \\
        R^{(3)} &= 2 Y^\circ \int_{[0,\eta)} \frac{\mathbb{E}[Y \mid X^u]}{\mathbb{P}(C > u \mid X)} \left(\diff 1_{(C \leq u)} - 1_{(C \geq u)} \frac{\mathbb{P}(C \in \diff u \mid X)}{\mathbb{P}(C \geq u \mid X)} \right).
    \end{align*}
    Straigtforward calculations give
    \begin{align*}
        \mathbb{E}[R^{(1)} \mid W] &= \mathbb{E}\left[\frac{Y^2}{\mathbb{P}(C \geq \eta \mid X)} \mid W \right], \\
        \mathbb{E}[R^{(3)} \mid W] &= -2 \mathbb{E}\left[Y \int_{[0,\eta)} \frac{\mathbb{E}[Y \mid X^u]}{\mathbb{P}(C > u \mid X)} \frac{\mathbb{P}(C \in \diff u \mid X)}{\mathbb{P}(C \geq u \mid X)} \mid W \right]. 
    \end{align*}
    Expanding the square gives $R^{(2)}=R^{(2.1)}+R^{(2.2)}+R^{(2.3)}$ for
    \begin{align*}
        R^{(2.1)} &= \left(\int_{[0,\eta)} \frac{\mathbb{E}[Y \mid X^u]}{\mathbb{P}(C > u \mid X)} \diff 1_{(C \leq u)}  \right)^2, \\
        R^{(2.2)} &= \left(\int_{[0,\eta)} \frac{\mathbb{E}[Y \mid X^u]}{\mathbb{P}(C > u \mid X)}  1_{(C \geq u)} \frac{\mathbb{P}(C \in \diff u \mid X)}{\mathbb{P}(C \geq u \mid X)} \right)^2, \\
        R^{(2.3)} &= -2 \int_{[0,\eta)} \frac{\mathbb{E}[Y \mid X^u]}{\mathbb{P}(C > u \mid X)} \diff 1_{(C \leq u)} \times \int_{[0,\eta)} \frac{\mathbb{E}[Y \mid X^u]}{\mathbb{P}(C > u \mid X)}  1_{(C \geq u)} \frac{\mathbb{P}(C \in \diff u \mid X)}{\mathbb{P}(C \geq u \mid X)}.
    \end{align*}
    Note  
    \begin{align*}
        &\mathbb{E}[R^{(2.2)} \mid W] \\
        &= \mathbb{E}\left[ \int_{[0,\eta)^2} \frac{\mathbb{E}[Y \mid X^u]}{\mathbb{P}(C > u \mid X)} \frac{\mathbb{E}[Y \mid X^v]}{\mathbb{P}(C > v \mid X)} 1_{(C \geq u \vee v)} \frac{\mathbb{P}(C \in \diff u \mid X)}{\mathbb{P}(C \geq u \mid X)} \frac{\mathbb{P}(C \in \diff v \mid X)}{\mathbb{P}(C \geq v \mid X)} \mid W \right] 
    \end{align*}
    by Fubini's theorem. By symmetry, this is twice the contribution where the indicator $1_{(C \geq v)}$ is replaced by $1_{(u \geq v)}$. Inserting this and towering on $X$ gives
    \begin{align*}
        &\mathbb{E}[R^{(2.2)} \mid W] \\
        &= 2 \mathbb{E}\left[ \int_{[0,\eta)} \frac{\mathbb{E}[Y \mid X^u]}{\mathbb{P}(C > u \mid X)} \left( \int_{[0,u]} \frac{\mathbb{E}[Y \mid X^v]}{\mathbb{P}(C > v \mid X)} \frac{\mathbb{P}(C \in \diff v \mid X)}{\mathbb{P}(C \geq v \mid X)}  \right) \mathbb{P}(C \in \diff u \mid X) \mid W \right].
    \end{align*}
    Straightforward calculations thus imply $\mathbb{E}[R^{(2.2)} \mid W] = -\mathbb{E}[R^{(2.3)} \mid W]$ so these terms cancel. Finally, note
    \begin{align*}
        \mathbb{E}[R^{(2.1)} \mid W] &= \mathbb{E}\left[ \int_{[0,\eta)} \frac{\mathbb{E}[Y \mid X^u]^2}{\mathbb{P}(C > u \mid X)^2} \mathbb{P}(C \in \diff u \mid X) \mid W \right]
    \end{align*} 
    Collecting the results gives
    \begin{align*}
        \Var[Y^\ast \mid W] &= \mathbb{E}\bigg[\frac{Y^2}{\mathbb{P}(C \geq \eta \mid X)} +\int_{[0,\eta)} \frac{\mathbb{E}[Y \mid X^u]^2}{\mathbb{P}(C > u \mid X)^2} \mathbb{P}(C \in \diff u \mid X) \\
        & \qquad \qquad -2 Y \int_{[0,\eta)} \frac{\mathbb{E}[Y \mid X^u]}{\mathbb{P}(C > u \mid X)} \frac{\mathbb{P}(C \in \diff u \mid X)}{\mathbb{P}(C \geq u \mid X)} \mid W \bigg] - \mathbb{E}[Y \mid W]^2.
    \end{align*}
   Note that $Y^2/\mathbb{P}(C \geq \eta \mid X)=\int_{[0,\eta)} Y^2 \diff \left( 1/\mathbb{P}(C > u \mid X) \right)+Y^2$
   and integration by parts implies
   $$\diff \left( \frac{1}{\mathbb{P}(C > u \mid X)} \right) = \frac{\mathbb{P}(C \in \diff u \mid X)}{\mathbb{P}(C \geq u \mid X)\mathbb{P}(C > u \mid X)}.$$
    Inserting this and collecting the integral terms implies
    \begin{align*}
        &\Var[Y^\ast \mid W] = \Var[Y \mid W] \\
        &+\mathbb{E}\bigg[ \int_{[0,\eta)} \bigg( \frac{Y^2}{\mathbb{P}(C \geq u \mid X)} + \frac{\mathbb{E}[Y\mid X^u]^2}{\mathbb{P}(C > u \mid X)} -\frac{2 \mathbb{E}[Y\mid X^u]^2}{\mathbb{P}(C \geq u \mid X)} \bigg) \frac{\mathbb{P}(C \in \diff u \mid X)}{\mathbb{P}(C > u \mid X)} \mid W\bigg].
    \end{align*}
    By writing $\mathbb{P}(C \in \diff u \mid X) = r(u \mid X) \diff \mu(u)$, one may take the expectation inside the integral and can then tower on $X^u$ and then take the expectation outside the integral again, leading to $Y^2$ being replaced by $\mathbb{E}[Y^2 \mid X^u]$. 
    By bounding $\mathbb{E}[Y\mid X^u]^2/\mathbb{P}(C > u \mid X) \geq \mathbb{E}[Y\mid X^u]^2/\mathbb{P}(C \geq u \mid X)$ and using Jensen's inequality for conditional expectations to bound $\mathbb{E}[Y^2 \mid X^u] \geq \mathbb{E}[Y \mid X^u]^2$ gives the desired conclusion.
\end{proof}

 \noindent Similarly to Theorem 3.1 in~\citet{Steingrimsson.etal:2019}, one could further have shown that $\Var[Y^\ast_{\mathbb{P},\mathbb{P}_2}(C,X^C) \mid W] \geq \Var[Y^\ast \mid W]$ so using a misspecified outcome distribution leads to larger variance of the pseudo-outcomes. This result is however not directly useful for our purposes and is hence omitted.

\section{Proof of Proposition~\ref{prop:asymp}} \label{sec:AppendixProof2}

\begin{proof}
    Write
    $$\hat{m}(w) - m(w) = \hat{m}(w) - \Tilde{m}(w) + \Tilde{m}(w) - m(w).$$
   To show stability of linear smoothers, note that
   \begin{align*}
        d_{w,D^{2n}}(0,\Var[Y^\ast(C,X^C) \mid W= \bigcdot \:]) &= \sum_{i=1}^n \left\{ \frac{p_i(w;W^n)^2}{\sum_{j=1}^n p_j(w;W^n)^2} \Var[Y^\ast(C,X^C) \mid W=W_i]^2  \right\} \\
        & \geq \inf_{z \in \{W_1,\dots,W_n\}} \Var[Y^\ast(C,X^C) \mid W=z]^2 \\
        & \geq \inf_{z} \Var[Y(X) \mid W=z]^2.
   \end{align*}
    where the last inequality follows from the second part of Theorem~\ref{thm:DRCUT}. This implies that $d_{w,D^{2n}}(0,\Var[Y^\ast(C,X^C) \mid W= \bigcdot \:])^{-1}$ is bounded and therefore also $O_{\mathbb{P}}(1)$. This result combined with condition (ii) gives stability of the linear smoother. Therefore $\hat{m}(w)-\Tilde{m}(w) = \hat{\mathbb{E}}_n[\hat{b}(W ; D^{n}_{I}) \mid D^{n}_{I}, W=w] + o_{\mathbb{P}}(n^{-\alpha})$.
   
   Introduce the stochastic norm
   \begin{align*}
       \lVert f(u,X;D^{n}_{I}) \rVert_{3,z, D^{n}_{I}} &= \left(\int_{[0,\eta)} \lVert f(u,X;D^{n}_{I}) \rVert_{z,D^{n}_{I}}^2 \diff \mu(u) \right)^{1/2}.
   \end{align*}
   By the first part of Theorem~\ref{thm:DRCUT}, 
    \begin{align*}
        \hat{b}(z;D^{n}_{I}) &= \int_{\mathcal{X} \times [0,\eta)} \left(\mathbb{E}[Y(X) \mid x^u] - \hat{\mathbb{E}}_{2,n}[Y(X) \mid x^u]\right)(\hat{\gamma}_{1,n}(u \mid x)-\gamma(u \mid x)) \\
        & \qquad \qquad \quad \frac{\mathbb{P}(C \geq u \mid x)}{\hat{\mathbb{P}}_{1,n}(C > u \mid x)} \mathbb{P}(X \in \diff x \mid W=z) \otimes \diff \mu(u) 
    \end{align*}
    so
   \begin{align*}
       \vert \hat{b}(z ; D^{n}_{I}) \vert \leq \varepsilon^{-1} \lVert \mathbb{E}[Y(X) \mid X^u] - \hat{\mathbb{E}}_{2,n}[Y(X) \mid X^u]  \rVert_{3,z,D^{n}_{I}} \lVert \hat{\gamma}_{1,n}(u \mid X)-\gamma(u \mid X)  \rVert_{3,z,D^{n}_{I}}
   \end{align*}
   by taking the absolute value onto the integrand, using positivity, and then employing the Cauchy-Schwarz inequality. Note
    \begin{align*}
        &\big\lvert \hat{\mathbb{E}}_n[\hat{b}(W ; D^{n}_{I}) \mid D^{n}_{I}, W=w] \big\rvert \\
        & \leq \sum_{i=1}^n \vert p_i(w;W^n) \vert \times \vert \hat{b}(W_{i} ; D^{n}_{I}) \vert\\
        &\leq \varepsilon^{-1} \sum_{i=1}^n  \vert p_i(w;W^n) \vert^{1/2} \lVert \mathbb{E}[Y(X) \mid X^u] - \hat{\mathbb{E}}_{2,n}[Y(X) \mid X^u]  \rVert_{3,W_i,D^{n}_{I}} \\
        & \qquad \qquad \quad \vert p_i(w;W^n) \vert^{1/2} \lVert \hat{\gamma}_{1,n}(u \mid X)-\gamma(u \mid X)  \rVert_{3,W_i,D^{n}_{I}}.
    \end{align*}
 By the Cauchy-Schwarz inequality
    \begin{align*}
        &\big\lvert \hat{\mathbb{E}}_n[\hat{b}(W ; D^{n}_{I}) \mid D^{n}_{I}, W=w] \big\rvert \\
        & \leq \varepsilon^{-1}  \left\{ \sum_{i=1}^n   \vert p_i(w;W^n) \vert \times \lVert \mathbb{E}[Y(X) \mid X^u] - \hat{\mathbb{E}}_{2,n}[Y(X) \mid X^u]  \rVert_{3,W_i,D^{n}_{I}}^2 \right\}^{1/2} \\
        & \qquad \times \left\{ \sum_{i=1}^n  
 \vert p_i(w;W^n) \vert \times \lVert \hat{\gamma}_{1,n}(u \mid X)-\gamma(u \mid X)  \rVert_{3,W_i,D^{n}_{I}}^{2} \right\}^{1/2} \\
 &= \sum_{i=1}^n \frac{\vert p_i(w;W^n) \vert}{\varepsilon} \times \lVert \mathbb{E}[Y(X) \mid X^u] - \hat{\mathbb{E}}_{2,n}[Y(X) \mid X^u]  \rVert_{2,w,D^{2n}} \\
 & \qquad \; \; \times \lVert \hat{\gamma}_{1,n}(u \mid X)-\gamma(u \mid X)  \rVert_{2,w,D^{2n}}
    \end{align*}
    where the final equality follows from the definition of the norm.
   
   Note that the sum of the absolute weights is $O_{\mathbb{P}}(1)$ by (iii). By Assumption (iv) and (v), the right hand side is thus $O_{\mathbb{P}}(n^{-\alpha_1-\alpha_2})$. To obtain the oracle rate, this term should be $o_{\mathbb{P}}(n^{-\alpha})$. This is satisfied if $n^{\alpha_1+\alpha_2} > n^{\alpha}$ or equivalently $\alpha_1 + \alpha_2 > \alpha$, which holds by (vi). It thus holds that $\hat{m}(w)-\Tilde{m}(w) = o_{\mathbb{P}}(n^{-\alpha})$ which implies $\hat{m}(w)-m(w) = O_{\mathbb{P}}(n^{-\alpha})$ as desired. 
\end{proof}

\section{Proof of Proposition~\ref{prop:crossFit}} \label{sec:AppendixProof3}

\begin{proof}
Note that
\begin{align*}
    \hat{m}^{\textnormal{CF}}(w) - m(w) &= \frac{1}{K}\sum_{k=1}^K (\tilde{m}_k(w)-m(w))  +  \frac{1}{K}\sum_{k=1}^K ( \hat{m}_k(w) - \tilde{m}_k(w) ). 
\end{align*}
Each $\tilde{m}_k(w)-m(w)$ can be analyzed analogously to $\tilde{m}(w)-m(w)$ but just using $n/K$ observations instead of $n$. Therefore $n^{\alpha}\{ \tilde{m}_k(w)-m(w) \} \rightarrow \mathcal{N}(K^\alpha \mu,K^{2\alpha} \sigma^2)$ in distribution since $n^\alpha = K^\alpha (n/K)^\alpha$. Furthermore, since $\tilde{m}_k(w)-m(w)$ for different values of $k$ are independent, one obtains $n^{\alpha}\{ 
K^{-1}\sum_{k=1}^K (\tilde{m}_k(w)-m(w)) \} \rightarrow \mathcal{N}(K^\alpha \mu,K^{2 \alpha - 1}\sigma^2)$ in distribution. For the second sum, note that each term $\hat{m}_k(w) - \tilde{m}_k(w)$ can be analyzed analogously to the sample split version $\hat{m}(w) - \tilde{m}(w)$. Under the assumptions from Proposition~\ref{prop:asymp}, it thus holds that $\hat{m}_k(w) - \tilde{m}_k(w) = o_{\mathbb{P}}(n^{-\alpha})$ so also $1/K \sum_{k=1}^K ( \hat{m}_k(w) - \tilde{m}_k(w) ) = o_{\mathbb{P}}(n^{-\alpha})$. Slutsky's lemma then implies, still under the assumptions from Proposition~\ref{prop:asymp}, that
\begin{align*}
    n^{\alpha}\{\hat{m}^{\textnormal{CF}}(w) - m(w)\} \rightarrow \mathcal{N}(K^\alpha \mu,K^{2\alpha-1} \sigma^2)
\end{align*}
in distribution. 
\end{proof}

\FloatBarrier